\documentclass[prb,aps,twocolumn,groupedaddress,floats,showpacs,final, fleqn]{revtex4-1}
\usepackage{dcolumn}
\usepackage{bm}
\usepackage{color}
\usepackage{mathtools}
\usepackage{amsmath,amsfonts,amssymb,bm,ulem}

\usepackage{graphicx}
\usepackage{epsfig}
\usepackage{psfrag}
\usepackage[pdftex]{hyperref} 

\begin{document}

\author{Olga Goulko}
\affiliation{Department of Physics, University of Massachusetts, Amherst, MA 01003, USA}
\author{Andrey S. Mishchenko}
\affiliation{RIKEN Center for Emergent Matter Science (CEMS),
2-1 Hirosawa, Wako, Saitama, 351-0198, Japan}
\affiliation{National Research Center ``Kurchatov Institute,"
123182 Moscow, Russia}
\author{Lode Pollet}
\affiliation{Department of Physics, Arnold Sommerfeld Center for Theoretical Physics, University of Munich, Theresienstrasse 37, 80333 Munich, Germany}
\author{Nikolay Prokof'ev}
\affiliation{Department of Physics, University of Massachusetts, Amherst, MA 01003, USA}
\affiliation{Department of Physics, Arnold Sommerfeld Center for Theoretical Physics, University of Munich, Theresienstrasse 37, 80333 Munich, Germany}
\affiliation{National Research Center ``Kurchatov Institute,"
123182 Moscow, Russia}
\author{Boris Svistunov}
\affiliation{Department of Physics, University of Massachusetts,
Amherst, MA 01003, USA}
\affiliation{National Research Center ``Kurchatov Institute,"
123182 Moscow, Russia}
\affiliation{Wilczek Quantum Center, Zhejiang University of Technology, Hangzhou 310014, China}

\title{Numerical analytic continuation: answers to well-posed questions}
\date{\today}

\begin{abstract}

We formulate the problem of numerical analytic continuation in a way that lets us draw meaningful conclusions about properties of the spectral function based solely on the input data. Apart from ensuring consistency with the input data (within their error bars) and the {\it a priori} and {\it a posteriori} (conditional) constraints, it is crucial to reliably characterize the accuracy---or even ambiguity---of the output. We explain how these challenges can be met with two approaches: stochastic optimization with consistent constraints and the modified maximum entropy method. We perform illustrative tests for spectra with a double-peak structure, where we critically examine which spectral properties are accessible and which ones are lost. For an important practical example, we apply our protocol to the Fermi polaron problem.
\end{abstract}


\maketitle

\section{Introduction}
\label{sec:I}

Numerous problems in science, from spectral analysis to image processing, require that we restore properties of a function $A(z)$ from a set of integrals
\begin{equation}
g_n = G[n,A] \equiv \int_{-\infty}^{\infty}\! dz K(n,z)  A(z), \; \; n=1, \dots , N,
\label{eq:ac0}
\end{equation}
where $K(n,z)$ is a known kernel and $\{ g_n \}$ is a finite set of experimental or numerical input data with error bars. An important class of such problems---known as numerical analytic continuation (NAC)---deals with ``pathological" kernels featuring numerous eigenfunctions with anomalously small eigenvalues. An archetypal NAC problem is the numerical spectral analysis at zero temperature, where the challenge is to restore the non-negative spectral function $A(z \ge 0)$ satisfying the equation
\begin{equation}
g_n = \int_{0}^{\infty}\! dz e^{-z \tau_n}  A(z) ,
\label{eq:spectral}
\end{equation}
from numerical data for $g_n=g(\tau_n \ge 0)$.

The NAC problem is often characterized as {\it ill-posed}. Mathematically, the near-degeneracy of the kernel implies two closely related circumstances:
(i) the absence of the resolvent, and
(ii) a continuum of solutions satisfying the input data within their error bars (even when integrals over $z$ are replaced with finite sums containing less or
equal to $N$ terms). Nowadays, the first circumstance is merely a minor technical problem, as there exists a number of methods allowing one to find solutions
to (\ref{eq:ac0}) without compromising the error bars of $g_n$.

The second circumstance---the ambiguity of the solution---is a more essential problem. It is clear that if one formulates the goal as to restore $A(z)$ as a continuous curve, or to determine its value on a given grid of points, then the goal cannot be reached as stated, irrespective of the properties of the kernel $K(n,z)$. The input data set is finite and noisy, thereby introducing a natural limit on the resolution of fine structures in $A(z)$.

Fortunately, the above-formulated goal has little to do with the practical world. In an experiment, all devices are characterized by a finite resolution function and the data they collect always correspond to {\it integrals}. The data are processed by making certain {\it assumptions} about the underlying function. This motivates an alternative formulation of the NAC goal involving integrals of $A(z)$ that render the problem well-defined. With additional assumptions about the smoothness and other properties of $A(z)$ behind these integrals, consistent with both {\it a priori} and {\it a posteriori} knowledge, the ambiguity of the solution can be substantially suppressed. The simplest setup is as follows:
\\

\noindent
{\it Given a set of finite intervals $\{ \Delta_m \}$, determine the integrals of
the spectral function over these intervals:
\begin{equation}
i_m = \Delta_m^{-1} \int_{z\in \Delta_m} dz A(z), \qquad m=1, \dots , M, \quad
\label{eq:ac1}
\end{equation}
along with the corresponding dispersions of fluctuations $\{ \sigma_m \}$ (straightforwardly extendable to the dispersion correlation matrix $\{ \sigma_{mm'} \}$).}
\\

\noindent Naively one might think that nothing is achieved by going from the integrals in (\ref{eq:ac0}) to the integrals in (\ref{eq:ac1}) because the latter have exactly the same form with the kernel $\bar{K}(m,z)=\Delta_m^{-1}$ for $z\in \Delta_m$ and zero otherwise (other forms of the ``resolution function" $\bar{K}(m,z)$ are discussed in Sec.~\ref{sec:SOCC}):
\begin{equation}
i_m = I[m,A] = \int_{-\infty}^{\infty} \! dz \bar{K}(m,z) A(z), \qquad m=1,\dots , M. \quad
\label{eq:ac2}
\end{equation}
This impression, however, is false because kernel properties are at the heart of the problem. If for appropriately small intervals (sufficiently small to resolve the variations of $A(z)$), the uncertainties for $i_m$ remain small, then one can draw reliable conclusions for the underlying behavior of $A(z)$ itself. The difference between ``good" (e.g. as in Fourier transforms) and pathological kernels is that for the latter, due to the notorious saw-tooth instability, the uncertainties for $i_m$ quickly become too large for a meaningful analysis of fine structures in $A(z)$.

To obtain a solution from the integrals (\ref{eq:ac1}), one has to invoke the notion of {\it conditional knowledge}. The most straightforward approach is to set the spectral function values at the middle points $z_m$ of the intervals $\Delta_m$ to $A_{\rm fin}(z_m)=i_m$. This is only possible if the intervals can be made appropriately narrow without losing accuracy for the integrals. With this approach we assume that the function is nearly linear over the intervals in question. This is a typical procedure for experimental data. Quantifying the error bar on $A_{\rm fin}(z_m)$ necessarily involves {\it two} numbers: the ``vertical'' dispersion $\sigma_m$ is directly inherited from $i_m$, and the ``horizontal"  error bar $\Delta_m/2$ represents the interval half-width.

\begin{figure}[htb]
\includegraphics[scale=0.38,width=0.97\columnwidth]{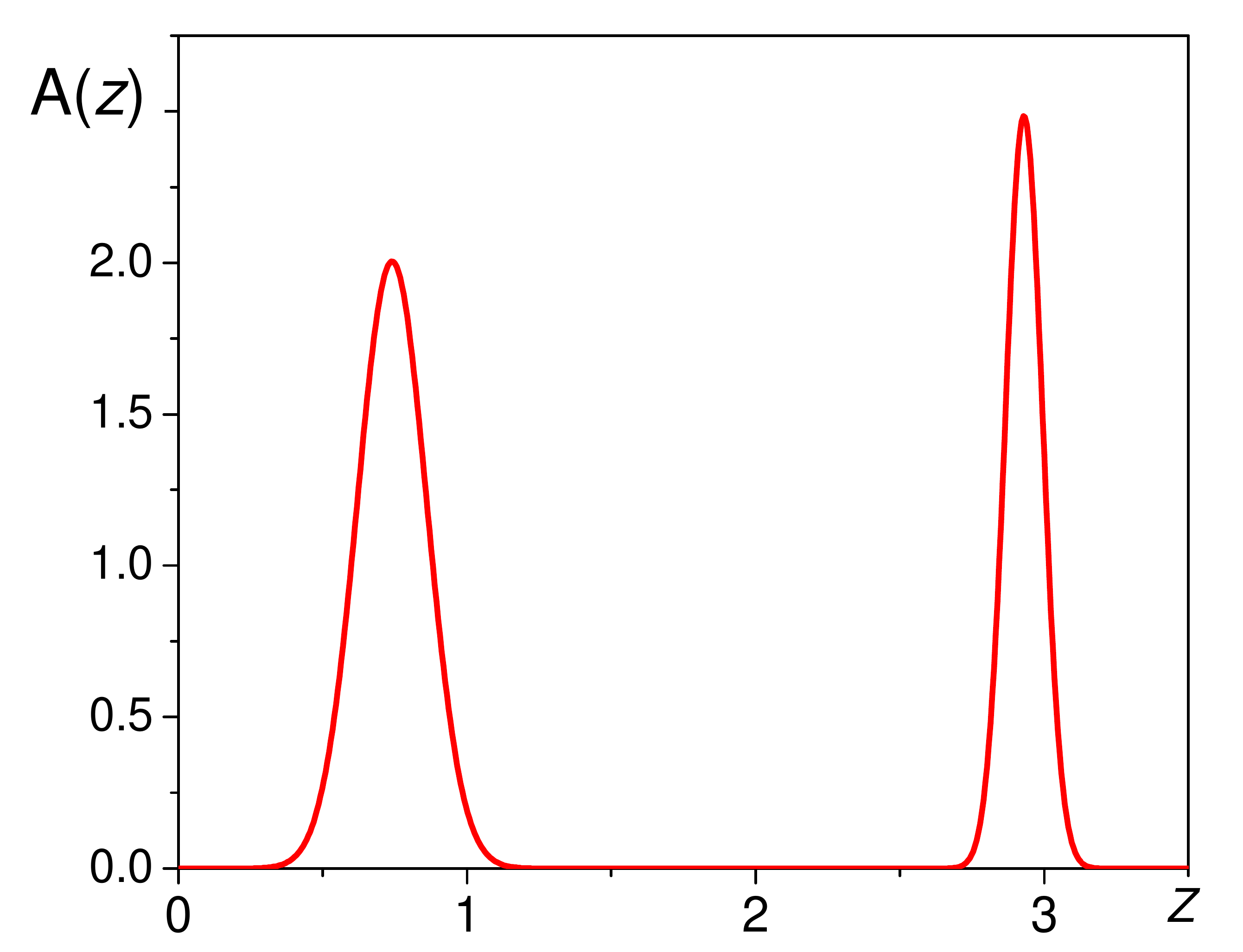}
\caption{Challenging example of a spectrum $A(z)$ for the NAC problem (\ref{eq:spectral}). As shown in the main part of the text,
the significant width of the first peak  makes it  essentially impossible to controllably restore the width of the second peak,
even with small relative error bars ($\sim 10^{-5}$) on $g_n$. On the other hand, the first two moments of the second peak, characterizing its weight and
position, can be extracted reliably.
}
\label{fig:two_peaks}
\end{figure}

The reader should be aware of two issues regarding such error bars. First, the error bars for different points are not independent but contain significant multi-point correlations. For example, an unrestricted integral $\int dz A(z)$ is typically known with an accuracy that is orders of magnitude better than what would be predicted by the central limit theorem if this integral is represented by a finite sum of integrals over nonoverlapping intervals. Second, the errors are not necessarily distributed as a Gaussian. Atypical fluctuations can have a significant probability and their analysis should not be avoided as the actual physical solution may well be one of them. To this end, it is important to explore the minimal and maximal values that the integral $i_m$ can take, and check that these are not significantly different from the typical value of $i_m$. In certain cases this criterion cannot be met without increasing the intervals $\Delta_m$ to an extent when the assumption of linearity of $A(z)$ becomes uncontrolled, implying that an important piece of information about the shape of $A(z)$ in this interval is missing. A characteristic example that plays a key role in the subsequent discussion is presented in Fig.~\ref{fig:two_peaks}, where the challenge is to extract the shape of the second peak.

In the more sophisticated approach used in this work, the values of $A_{\rm fin}(z_m)$ can be further optimized (without compromising the accuracy of the solution with respect to $g_n$) to produce a smooth curve. This protocol has the additional advantage of eliminating minima, maxima, gaps, and sharp features that are not guaranteed to exist by the quality of input data. The nature of the problem is such that very narrow peaks (or gaps) with tiny spectral weight can always be imagined to be present (for narrow intervals they will certainly emerge due to the saw-tooth instability). Our philosophy with respect to these features is to erase them {\it within the established error bounds} and obtain a solution that is insensitive to the interval parameters.

Having established a smooth solution, one may nevertheless ask whether a particular feature of the solution can, in principle, have significantly different properties. For example, if the NAC procedure suggests a peak, one may wonder if the true spectral function could have a much narrower peak with the same area, and if so, what is its smallest possible width. A NAC protocol should be able to answer this type of question fast and reliably. In this work, we explain how these goals can be achieved in practice. Many technical details of the protocol that we propose to abbreviate as SOCC (Stochastic Optimization with Consistent Constraints) were already published in Refs.~\onlinecite{SOM,Julich,CC} as separate developments.
The crucial advances here are (i) the final formulation based on integrals of the spectral function, and (ii) the idea of working with linear combinations of pre-calculated ``basic" solutions. The latter allows one to readily apply consistent constraints without compromising the error bars on the input data. Consistent constraints are also crucial for assessing what features can be resolved and what information is unrecoverable.

In what follows, the term ``consistent constraints" applies to (i) the general principle of utilizing the  {\it a priori}  and revealing the  {\it a posteriori} (conditional) knowledge without compromising the error bars of the input data and (ii) a particular set of numerical procedures based on ideas respecting this principle. Our SOCC protocol involves two different consistent-constraint procedures. The first one, borrowed form
the NAC method of Ref.~\onlinecite{CC}, is now used solely to (dramatically!) enhance the performance of the stochastic-optimization part of the protocol searching for basic solutions.
The most important consistent-constraint procedure is used to post-process the set of basic solutions.

The paper is organized as follows. In Sec.~\ref{sec:SOCC} we describe the SOCC method consisting of three distinct stages, and explain how a smooth solution can be obtained without any bias with respect to solving Eq.~(\ref{eq:ac0}) and analyzed for possible atypical deformations. In Sec.~\ref{sec:MaxEnt} we briefly review the maximum entropy method (MEM).\cite{MaxEnt1,MaxEnt1a,MaxEnt1b,MaxEnt2,MaxEnt3,MaxEnt5} In Sec.~\ref{sec:tests} we explore what SOCC and MEM methods predict for the test spectral function shown in Fig.~\ref{fig:two_peaks}, and how one should analyze the final solution with respect to its possible smooth transformations. In Sec.~\ref{sec:polarons} we apply our findings to the physical spectral function of the resonant Fermi polaron.\cite{polaronexp1, grimmexprepulsive, polaronexp3, new, parish, Kamikado, polaronfrg, repulsivepolaron} We conclude in Sec.~\ref{sec:conclusions}.

\section{Stochastic Optimization with Consistent Constraints}
\label{sec:SOCC}

The formulation of the SOCC method is relatively simple and consists of three parts: \\

1.~Finding a large set of solutions $A_j(z)$ [$j=1, \dots , J\gg 1$]
to Eq.~(\ref{eq:ac0}) that satisfy the input data within their
error bounds. In what follows we call them ``basic'' solutions.
Basic solutions are not biased in any way to be smooth or to satisfy
any other requirements based on knowledge about the problem
outside of Eq.~(\ref{eq:ac0}). The irregularity of basic solutions embodies what is referred to as an ill-posed problem. In subsection \ref{subsec:SO} we briefly explain
how these solutions are found by the stochastic optimization procedure
(most technical details were published previously in Refs.~\onlinecite{SOM,Julich})
and how the consistent constraints method \cite{CC} is used to improve drastically the speed of the stochastic
optimization protocol.
\\

2.~Using the basic solutions $A_j(z)$ to compute the integrals~(\ref{eq:ac2}) with a different kernel
$\bar{K}(m,z)$. There are several choices here. One of them is given by Eq.~(\ref{eq:ac1})
and amounts to computing integrals from $ A_j(z) $ over finite intervals $\{ \Delta_m \}$
centered at $\{ z_m \}$. However, one is also free to consider normalized continuous
kernels with unrestricted integration over $z$, such as Lorentzian (or Gaussian) shapes
centered at points $\{ z_m \}$ with the width $\{ \Delta_m \}$ at half-height, e.g.,
\begin{equation}
\bar{K}(m,z) =\frac{\Delta_m/\pi}{(z-z_m)^2 +\Delta_m^2} \;.
\label{eq:Lorentz}
\end{equation}
Thus obtained sets of integrals $\{ i_{m}^{(j)} \}$ are then used straightforwardly
to compute averages
\begin{equation}
i_m = J^{-1} \sum_{j=1}^{J} i_{m}^{(j)} \;,
\label{eq:av}
\end{equation}
and dispersions \cite{Maier_note}
\begin{equation}
\sigma_m^2 = J^{-1} \sum_{j=1}^{J} \left( i_{m}^{(j)} - i_m \right)^2 \;.
\label{eq:disp}
\end{equation}
To characterize possible two-point correlations, one should compute
the correlation matrix
\begin{equation}
\sigma_{m m'} = J^{-1} \sum_{j=1}^{J}
\left( i_{m}^{(j)} - i_m \right) \left( i_{m'}^{(j)} - i_{m'} \right) \;.
\label{eq:corr}
\end{equation}
Strictly speaking, there is no reason to stop characterizing correlations
at the two-point level. One may proceed with computing multi-point averages but the effort
quickly becomes expensive and the outcome cannot be presented in a single plot. An alternative ``visualization" of multi-point correlations is discussed in subsection \ref{subsec:final}. \\

3.~Interpreting the result. The simplest interpretation and an estimate of the dispersion
for typical fluctuations is to assume that $A(z)$ is nearly linear over the range of each interval. This leads to the solution $A_{\rm fin}(z_m) = i_m$ with vertical and horizontal
``error bars" $\sigma_m^{(v)}=\sigma_m$ and $\sigma_m^{(h)}=\Delta_m/2$.
The vertical error bars may be overestimated because fluctuations at different points
are correlated. However, as explained in the Introduction, the correct answer may
correspond to some atypical shape, and this possibility has to be addressed as well.

An alternative protocol, discussed in subsection \ref{subsec:final}, determines the final solution
by selecting a superposition of basic solutions
\begin{equation}
A_{\rm fin}(z) = \sum_{j=1}^{J} c_j A_j(z), \qquad \quad \sum_{j=1}^{J} c_j =1  \;,
\label{eq:final}
\end{equation}
such that $A(z)$ remains non-negative (with high accuracy) and the coefficients
$c_j$ are optimized to impose smooth behavior or any other ``conditional knowledge''. Formally, the simplest interpretation corresponds to $c_j=1/J$.

\subsection{Search for basic solutions}
\label{subsec:SO}

The search for basic solutions relies on the minimization of
\begin{equation}
\chi^2 = N^{-1}\sum_{n=1}^{N} \left( \frac{g_n-G[n,A]}{\delta_n}\right)^2 \;,
\label{eq:chi2}
\end{equation}
where $\delta_n$ is the error of the $g_n$ value. Without loss of generality,
we assume that the components of the vector $\vec{g}=(g_1,g_2, \dots )$ are uncorrelated;
otherwise, one has to perform a rotation to the eigenvector basis of the two-point
correlation matrix
$ \langle (g_n-\langle g_n \rangle ) (g_n'-\langle g_n' \rangle ) \rangle$
where the components of $\vec{g}$ become statistically independent.
This linear transformation leads to an equation that has exactly the same
form as (\ref{eq:ac0}) with a rotated kernel. We choose a maximal tolerance $\chi_c$ of order unity and search for functions $A(z)>0$ with $\chi^2 < \chi_c^2$, which are then added to the set of basic solutions for further processing.

\begin{figure}[htb]
\includegraphics[scale=0.38,width=0.86\columnwidth]{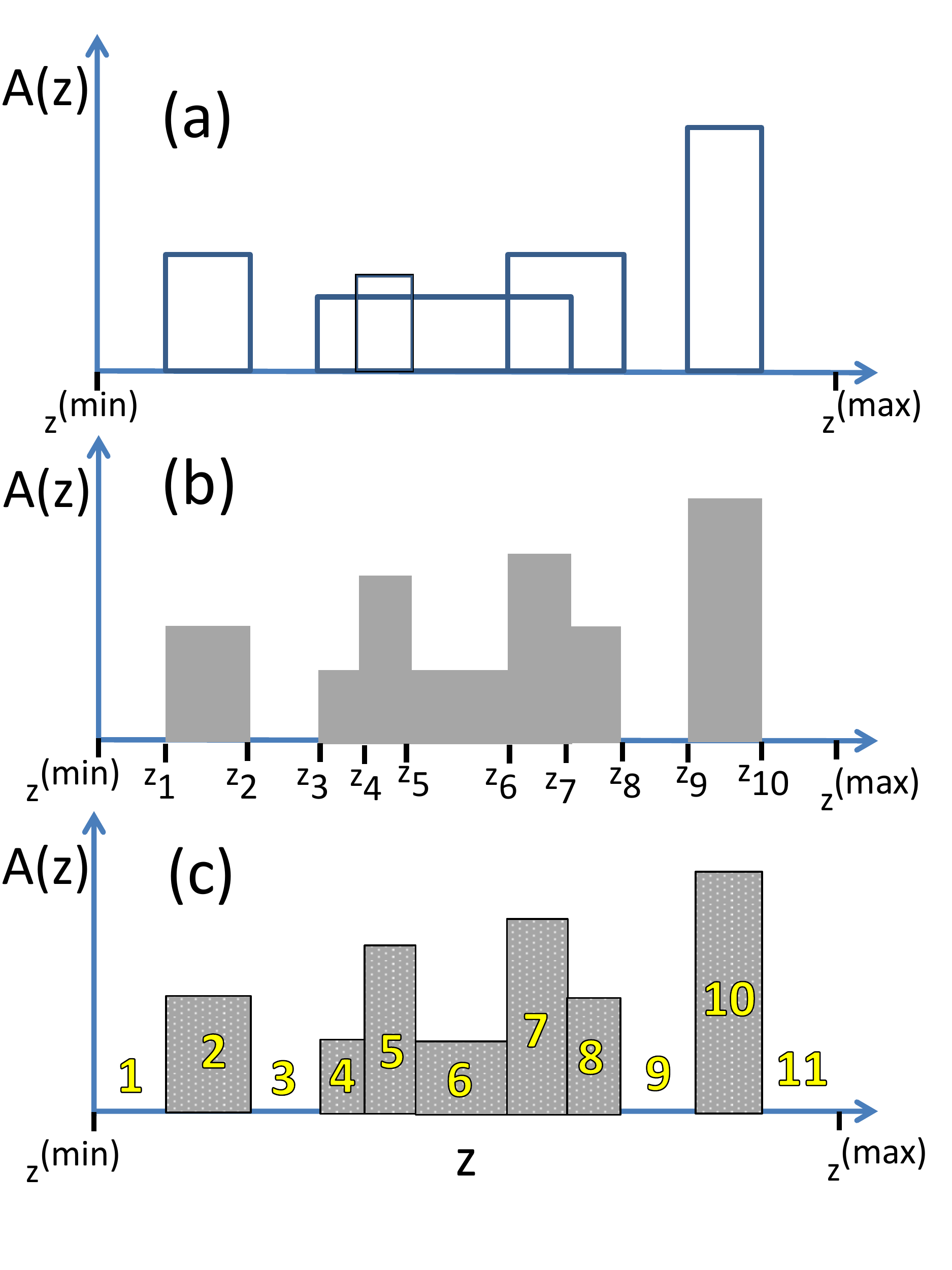}
\caption{ (a) Spectral function parametrization by
a set of rectangles. (b) Illustration of the treatment of intersections of rectangles. (c) Identical re-parametrization of the spectrum.}
\label{fig:Slicing}
\end{figure}

Information about the input data is limited to the objective function
(\ref{eq:chi2}). Truly unbiased methods should not assume anything about $A(z)$
that is not part of exact knowledge, such as the predetermined grid of points, and the number
and parameters of peaks/gaps.
In the stochastic optimization method of Refs.~\onlinecite{SOM,Julich}, each solution is represented
by a set of positive-definite rectangular shapes (the $\delta$-function can be viewed
as the limiting case of an infinitely narrow and infinitely high rectangular shape
with fixed area), which are allowed to have multiple overlaps, see panel (a) in Fig~\ref{fig:Slicing}.
More precisely, a solution is represented as a sum
\begin{equation}
A(z) = \sum_{r=1}^{R} \eta_{\{ P_r \}}(z)
\label{AA}
\end{equation}
of rectangles $\{ P_r \} =  \{ h_r,w_r,c_r \}$,
\begin{eqnarray}
\eta_{\{ P_r \}}(z) = \left\{
\begin{array}{ll}
h_r\, , & \quad  z \in [c_r-w_r/2,c_r+w_r/2] \, , \\
0\, , & \quad \mbox{otherwise}\, ,
\end{array}
\right.  \quad
\label{rectang}
\end{eqnarray}
where $h_r>0$, $w_r>0$, and $c_r$ are the height, width, and center
of the rectangle $P_r$, respectively. In what follows we refer to
\begin{equation}
{\cal C} = \left\{  {\{ P_r \}}, \, r=1, ... , R  \right\} \;,
\label{config}
\end{equation}
as a ``configuration.'' All rectangles are restricted to
the specified range of the $A(z)$ support,
$[z^{\textnormal{(min)}},z^{\textnormal{(max)}}]$; i.e.,
for all rectangles $c_r-w_r/2>z^{\textnormal{(min)}}$ and
$c_r+w_r/2<z^{\textnormal{(max)}}$.
The spectrum normalization is given by
$\sum_{r=1}^{R} h_r w_r = N_0$, and $G[n,A]$ in Eq.~(\ref{eq:ac0}) can be
written as
\begin{equation}
G[n,A]= \sum_{r=1}^{R} {\cal K}(n,r) h_r  \; ,
\label{simulated-discrete}
\end{equation}
where
\begin{equation}
{\cal K}(n,r) =
\int_{c_r-w_r/2}^{c_r+w_r/2} d z \; K(n,z) \; .
\label{I_m}
\end{equation}

The number of rectangles and all continuous parameters
characterizing their position, width, and height are found by minimizing the
objective (\ref{eq:chi2}). Optimization starts from a randomly generated set of rectangles,
and finds a large number of dissimilar basic solutions $ A_j(z)$
with $\chi^2<\chi_c^2$. More precisely, the search is based on a chain of randomly chosen
updates over the configuration space of shapes which fully explore the saw-tooth fluctuations
of basic solutions. This is important for the successful elimination of noise
in the final solution.

Updates proposing small modifications of the shape (``elementary'' updates) have the disadvantage of long computation time for a basic solution. To speed up the search, we supplement the standard
protocol of Refs.~\onlinecite{SOM,Julich} with consistent-constraints
(CC) updates, which propose a radical shape modification based on minimization of
the positive-definite quadratic form $\chi^2 + \sum_i o_i$ by matrix inversion
as described in Ref.~\onlinecite{CC}. Here $o_i$ are various positive-definite quadratic forms,
or ``penalties,'' that ensure that the matrix to be inverted is well-defined. This is achieved
by penalizing the  derivatives of $A(z)$ (computed on the grid based on the current
 configuration $\{ {\mathcal C} \}$, see below) and enforcing $A(z)\ge0$.
The CC-update involves a number of iterations when penalties $o_i$ are adjusted
self-consistently in such a way that at the end of the update $0< \sum_i o_i \ll \chi^2$.
Explicit forms for $\{ o_i \}$ and the adjustment protocols are described in detail in
Ref.~\onlinecite{CC} (see also $O_1$ and $O_4$ forms in subsection \ref{subsec:final}).

Even though the CC-updates do not compromise the goal of minimizing $\chi^2$, their efficiency is based on penalties that suppress saw-tooth fluctuations.
To exclude possible bias originating from CC-updates on basic solutions we proceed as follows.
The global update of the SOCC method consists of thousands of elementary
updates $L_{\textnormal{tot}}$ that are divided into two groups:
$L_a$ stage-a updates and $L_b$ stage-b updates, where
$L_a+L_b=L_{\textnormal{tot}}$ and $L_a<L_b$.
Updates increasing $\chi^2$ are temporarily considered ``accepted"
(and the resulting configuration recorded) with high probability during stage-a,
but this probability is reduced during stage-b that favors updates
decreasing $\chi^2$. The idea is to use $L_a$ updates
to escape from the local minimum of $\chi^2$ in the multi-dimensional
configuration space in a hope to find a better minimum afterwards.
The global update is accepted only if a smaller value of $\chi^2$ was recorded
in the course of applying elementary updates, and the new configuration
becomes the one with the smallest $\chi^2$. We apply CC-updates during stage-a
of a global update when the increase of $\chi^2$ is allowed, and proceed with
a large number of elementary updates, which results in a configuration with
fully developed saw-tooth instability.

We found that CC-updates have no effect on the self-averaging of the
saw-tooth noise in the equal-weight superposition of basic solutions,
improve typical $\chi^2$-values for basic solutions, and significantly
decrease the computation time required for finding basic solutions.

To run the CC-update, one has to re-parameterize the configuration as a collection of nonoverlapping rectangles in order to be able to
use their heights for estimates of the function derivatives.
Panel (b) in Fig.~\ref{fig:Slicing} illustrates how overlaps of
rectangles are understood in the SOCC method. This leads to an identical
re-parametrization in terms of nonoverlapping rectangles
$\{ \widetilde{P}_r \} =  \{ \widetilde{h}_r,\widetilde{w}_r,\widetilde{c}_r \}$.
The conversion is done as follows. First, the set of rectangle parameters
$\{ c_r-w_r/2 \} \cup \{ c_r+w_r/2 \}$ is ordered to form a grid of new
bin boundaries that also include the support limits $z^{\textnormal{(min)}}$ and $z^{\textnormal{(max)}}$. Second, bin centers and widths  become centers and widths
of the ordered set of new rectangles, respectively:
\begin{eqnarray}
&\widetilde{c}_{r+1} > \widetilde{c}_{r} \;\; \forall r \;, &
\label{cond1} \\
&\widetilde{c}_1-\widetilde{w}_1/2  = z^{\textnormal{(min)}},& \nonumber \\
&\widetilde{c}_r-\widetilde{w}_r/2  =
\widetilde{c}_{r-1} +\widetilde{w}_{r-1}/2 \;\; \forall r, & \label{cond2} \\
&\widetilde{c}_{2R+1}+\widetilde{w}_{2R+1}/2  = z^{\textnormal{(max)}} \; .
& \nonumber
\end{eqnarray}
Figures~\ref{fig:Slicing}(b) and ~\ref{fig:Slicing}(c) illustrate how the conversion from
$\{ P_t \}$ to $\{ \widetilde{P}_t \}$ amounts to an identical representation
of the spectrum: $R$ original rectangles introduce $2R$
boundaries on the $[z^{\textnormal{(min)}},z^{\textnormal{(max)}}]$ interval
and split it into $2R+1$ rectangles obeying conditions (\ref{cond1})--(\ref{cond2}).
Note that some rectangles have zero height when submitted into the CC-update.
The update modifies the values of all $\widetilde{h}$-parameters and generates
a new set $\{ \widetilde{h}_r' \}$. Since $\{ \widetilde{P}_r \}$
is a particular case of $\{ P_r \}$, there is no need to perform any
additional transformation to proceed with elementary updates.

\subsection{Preparing the final solution}
\label{subsec:final}

Because all basic solutions satisfy Eq.~(\ref{eq:ac0}), one can immediately
check that a linear combination of basic solutions, Eq.~(\ref{eq:final}),
always leads to a solution of Eq.~(\ref{eq:ac0}) with the same accuracy cutoff $\chi_c$
as the basic solutions, provided that all $c$-coefficients are non-negative.
Indeed, by linearity of the problem and the condition $\sum_{j=1}^{J} c_j = 1$,
\begin{eqnarray}
G[n,A_{\rm fin}] & = & \sum_{j=1}^{J} c_j G[n,A_j] \;, \nonumber \\
g_n-G[n,A_{\rm fin}] & = & \sum_{j=1}^{J} c_j \left( g_n-G[n,A_j] \right) \;.
\label{eq:lin1}
\end{eqnarray}
Substituting these expressions into the $\chi^2$ form for the final solution
and employing the Cauchy--Bunyakovsky--Schwarz inequality for $\chi_{jj'}$ we get
\begin{eqnarray}
{} && \chi^2  =  \sum_{j,j'=1}^{J} c_j c_{j'} \chi_{jj'} \le \bar{C}^2\,
\chi_c^2, \;\;\mbox{with} \;\; \bar{C}= \sum_{j=1}^{J} \vert c_j \vert , \label{eq:lin2a} \\
{} && \chi_{jj'} =  N^{-1}\sum_{n=1}^{N}
 \frac {(g_n-G[n,A_j] )(g_n-G[n,A_{j'}] )}{\delta_n^2} \;. \qquad \;\;
\label{eq:lin2}
\end{eqnarray}

If some $c$-coefficients are negative, the accuracy of the final solution
is guaranteed only if $\bar{C}$ is not large. One may argue that the upper bound
$ \chi^2  \le \bar{C}^2 \, \chi_c^2$ is substantially overestimating deviations,
and the actual accuracy is better. Let $C_{+}=(1+\bar{C})/2$ and $C_{-}=(1-\bar{C})/2$
be the sums over all positive and all negative coefficients, respectively.
Then linear superpositions of basic solutions involving only positive and
only negative coefficients and divided by
$C_{+}$ and $C_{-}$ (we denote them as $A_{+}$ and $A_{-}$, respectively), have their
$\chi^2$ measures smaller or equal to $\chi_c^2$, by Eq.~(\ref{eq:lin2a}).
The final solution can be identically written as $A_{\rm fin}=C_{+}A_{+} + C_{-}A_{-}$,
and its $\chi^2$-measure is nothing but the two state version of Eq.~(\ref{eq:lin2a}).
Since the $G[n,A]$ values are derived from spectral density integrals they are
smooth functions of $n$ and random point-to-point sign fluctuations of $g_n-G[n,A]$
are arising predominantly from $g_n$. Thus, the expectation is that $\chi_{+-}$ is
positive, in which case
\begin{equation}
\chi^2 \le (C_{+}^2 +C_{-}^2)\chi_{c}^2 - 2C_{+}\vert C_{-} \vert \chi_{+-}
\le \frac{1+\bar{C}^2}{2}\, \chi_c^2 .
\label{eq:lin2c}
\end{equation}
In practice, sign-positivity of the spectral density severely
restricts the possibility of having large $|C_{-}|$ and $\bar{C}$
in the final solution and $|C_{-}|$ tends to remain smaller than unity automatically.
Finally, Eq.~(\ref{eq:lin2a}) is only an upper bound, and superpositions
with $\bar{C}$ as large as $2$ may still have $\chi^2< \chi_c^2$.

These considerations lead to an important possibility of modifying the shape of the final
solution in order to satisfy additional criteria formulated outside of
Eq.~(\ref{eq:ac0}). The key observation, and crucial difference to
other NAC methods, is that ``conditional knowledge'' protocols are invoked
{\it after} all basic solutions are determined, meaning that they remain
unbiased with respect to the input data.

As discussed in the Introduction, the most conservative philosophy regarding
sharp spectral features, such as peaks and gaps, is to eliminate them if they
are not warranted by the quality of the input data. (Our method does allow
to answer the question whether a given sharp feature is compatible
with the input data, see below.) To implement the idea, we formulate the
problem of determining an appropriate set of $\{ c_j \}$ coefficients
as a linear self-consistent optimization problem closely following the
consistent constraints method of Ref.~\onlinecite{CC}. The objective function to
be minimized consists of several terms, $O=\sum_{k=1}^{5} O_k$, each being a
quadratic positive-definite form of $c_j$. More terms can be added if necessary
to control higher-order derivatives, enforce expected asymptotic behavior, etc.\\

$\bullet$ To suppress large derivatives we consider the following form
\begin{equation}
O_1= \sum_{k=2}^{K} \left\{ D_k^2 [A'(z_k)]^2 + B_k^2 [A''(z_k)]^2 \right\} \;,
\label{eq:O1}
\end{equation}
where $\{ z_k \}$ is the grid of points used to define
the first and second discrete derivatives of the function $A(z)$.
The sets of coefficients $D_k$ and $B_k$ are adjusted under iterations
self-consistently in such a way that contributions of all $z_k$-points to $O_1$ are similar.

$\bullet$ The unity-sum constraint on the sum of all coefficients in the superposition
is expressed as
\begin{equation}
O_2= {\cal U} \left( \sum_{j=1}^{J} c_j-1 \right)^2 \;,
\label{eq:O2}
\end{equation}
with a large constant ${\cal U}$.

$\bullet$ Since $O_1+O_2$ does not constrain the amplitudes and signs of all
$c_j$ the minimization cannot proceed by matrix inversion. To improve matrix
properties we add a ``soft'' penalty for large deviations of $c_j$ from the
equal-weight superposition
\begin{equation}
O_3= \sum_{j=1}^{J}\, (c_j-1/J)^2 \;.
\label{eq:O3}
\end{equation}

$\bullet$ To ensure that the spectral function is non-negative (with high accuracy)
we need $z$-dependent penalties (to be set self-consistently) that suppress the
development of large negative fluctuations:
\begin{equation}
O_4= \sum_{k=1}^{K}  Q_k A(z_k)^2 \;.
\label{eq:O4}
\end{equation}

$\bullet$  Finally, we can introduce a penalty for the solution to deviate from some ``target''
function (or ``default model'') $A_T(z_k)$:
\begin{equation}
O_5= \sum_{k=1}^{K} \, T_k  \, [A(z_k)-A_T(z_k)]^2 \;.
\label{eq:O5}
\end{equation}
The main purpose of $O_5$ is to address subtle multi-point correlations between allowed shapes: by forcing the solution to be close to a certain target function one can monitor how the solution starts developing additional saw-tooth-instability-related features or violates the unity-sum constraint. This penalty is zero when preparing $A_{\rm fin}$ in the absence of any target function.
\\

The self-consistent optimization protocol is as follows. We start with
$c_j=1/J$ and compute $A(z_k)$. The initial sets of coefficients in $O_1$
are defined as $D_k = {\cal D}$, and $B_k = {\cal D}$, where ${\cal D}$ is some
small positive constant (its initial value has no effect on the final solution
because  penalties for derivatives will be increased exponentially under iterations).
Since the positivity of $A(z)$ is guaranteed in the initial state, we set $Q_k=0$.
After the quadratic form for the objective function $O$ is minimized, the new set of
$c$-coefficients is used to define a new solution $A(z)$,
penalties for derivatives are increased, ${\cal D}\to f {\cal D}$, $D_k\to fD_k$,
$B_k\to fB_k$ by some factor $f>1$, and then all penalties in
in $O_1$ and $O_4$  are adjusted self-consistently as follows:
\begin{itemize}
\item  If $D_k |A'(z_k)| > {\cal D}$,  we assign $D_k = {\cal D}/|A'(z_k)|$;
\item  If $B_k |A''(z_k)| > {\cal D}$, we assign $B_k = {\cal D}/|A''(z_k)|$;
\item  If $A(z_k) < 0$,                we assign a large penalty suppressing the amplitude of the solution at this point, $Q_k = {\cal Q}$, where ${\cal Q}$ is a large constant;
otherwise the value of $Q_k$ is increased by two orders of magnitude.
\end{itemize}
In this work we use ${\cal U}=10^6$, ${\cal Q}=10^6$, and $f=2$. This sets the stage for the next iteration of the $O$-optimization protocol.

Since the accuracy expression, Eq.~(\ref{eq:lin1}), relies on the substitution $g_n = \sum_{j=1}^{J} c_j g_n$, it is crucial that
the unity-sum constraint is satisfied for all input data points,
\begin{equation}
 \left\vert \sum_{j=1}^{J} c_j -1 \right\vert < \epsilon \, , \qquad
\epsilon = \min \{ \vert \delta_n/g_n \vert \}_n \;.
\label{eq:stop}
\end{equation}
This provides the required criterion for terminating iterations. The final solution (\ref{eq:final}) is based on the last set of $c$-coefficients that satisfied the condition $\chi^2 < \chi_c^2$.

In the absence of the target objective $O_5$, the procedure is guaranteed to produce
a final solution $A_{\rm fin}(z)$ with smooth behavior because our initial solution
already satisfies all requirements. The derivative objective is forcing $A(z)$
to be as smooth as possible within the subspace of fluctuations that keep
$\chi^2$ small.

With the help of $O_5$ one can explore how the solution
is modified if one forces it to go through some set of points.\cite{CC} The simplest case
would be to set $T_k={\cal T}$ for some point $z_o$ (where ${\cal T}$ is a large number;
in this work ${\cal T}=10^6$) and zero otherwise, and shift $A_T(z_o)$ away from $A_{\rm fin}(z_0)$.
The solution going through the point  $A_T(z_o)$ is no longer guaranteed to be
smooth in the vicinity of $z_o$ and for large deviations from the final solution
will develop the saw-tooth instability at $z_o$.

The most interesting choices for $z_o$ are the minima and maxima of the spectrum.
Despite the fact that our protocol is to erase sharp features not warranted by
the input data quality, we can still address questions such as ``can this spectral peak
(or gap) be made narrower/higher/lower and by how much, before the solution
becomes unstable against developing secondary features?'' This question cannot be fully
answered at the level of the correlation matrix (\ref{eq:corr}) because
(i) spectral functions have subtle multi-point correlations, and
(ii) the notion of a ``typical" solution has no physical meaning in this context.
The only way to answer this question is to have access to a
large representative set of unbiased basic solutions.

The objective $O_5$ offers a generic way of exploring various possibilities for
underlying features hidden behind the accuracy of input data. Clearly, there are other
alternatives for addressing specific questions. For example, one can isolate a spectral
peak to some interval and compute the dispersion $d_j$ of each basic solution
$j$ over this interval. Next, the distribution function $W(d)$ over all basic solutions
is composed and analyzed. If $W(d)$ has a narrow region of support around its average
$\langle d \rangle$ value, then the peak width cannot significantly deviate from
$\langle d \rangle$. If $W(d)$ is nonzero for $d \ll \langle d \rangle$ then
one has to conclude that the actual peak might be much narrower (and, correspondingly,
have a much higher amplitude) than what is predicted by the typical smooth solution
$A_{\rm fin}$.

\section{Maximum Entropy method}
\label{sec:MaxEnt}

Numerous NAC schemes are based on a totally different philosophy and impose
additional restrictions/penalties on the allowed functional shapes of $A(z)$
{\it in the process} of solving Eq.~(\ref{eq:ac0}). In other words, the search
for solutions is biased with ``conditional knowledge''
from the very beginning. Historically, the Tikhonov-Phillips regularization method\cite{Ti1,Ph,Tikh-Arse,Tikh_etal}
was the first to advocate this approach.
Currently, the most popular scheme of this type is the maximum entropy
method.\cite{MaxEnt1,MaxEnt1a,MaxEnt1b,MaxEnt2,MaxEnt3,MaxEnt5}
Other schemes worth mentioning are singular value decomposition,\cite{svd}
non-negative least-squares,\cite{nnls} stochastic regularization,\cite{sr}
and averaging Pad{\'e} approximants.\cite{pade,pade2} In the stochastic sampling method
of Refs.~\onlinecite{Sand,Vafayi, Beach, MaxEnt3}, the remaining bias is in the form of
the predetermined grid of frequency points, and the final solution is an average
over a certain ``thermal'' ensemble (see also Ref.~\onlinecite{Sandvik2015} for a further refinement).
Fast effective modification of stochastic optimization (FESOM) \cite{Maier} also uses
the predetermined grid of frequency points.

We now briefly review the maximum entropy method,\cite{MaxEnt2} which can be seen
as a special case of stochastic sampling methods.\cite{Sand, Beach} Instead of minimizing $\chi^2$ one constructs a functional $Q = \frac{1}{2}\chi^2 - \alpha S[A]$,
where $S[A]$ is the ``entropy'' term. The positive parameter $\alpha$ is a Lagrange multiplier that can also be thought of as a ``temperature'' by analogy to classical statistics
(note that our definition of $\chi^2$ differs by a factor of $N$ from the standard MEM formulation; this is, however, only a matter of convention).
The entropy term $S[A]$ takes the form $S[A] = -\int dz A(z) \ln \left[ A(z) / M(z) \right]$ with $M(z)$ being the default model.
For very large values of $\alpha$, the default model term
dominates in $Q$, reflecting our ignorance about the system. For very low values
of $\alpha$, the ``energy'' term $\chi^2$ dominates, reflecting the quality of the input data.
For intermediate values of $\alpha$, one interpolates between these two limits
and obtains a trade-off between accuracy and smooth behavior enforced by the default model.
We are using Bryan's method~\cite{MaxEnt1a} to implement the minimization procedure: the final
answer is obtained by averaging over all values of $\alpha$ weighted with the respective {\it a posteriori} probability (we saw however little difference between Bryan's method and the classical MEM in the examples below). In Bryan's method, a singular value decomposition is also applied,
which reflects the fact that the finite precision of storing floating point numbers in combination with the poor conditioning of the kernel puts severe limitations on the information that can possibly be retrieved. One can hence reduce the search space at no substantial loss; in practice, only 5 to 20 search directions survive this step. The remaining minimization is performed by the
Levenberg-Marquadt algorithm. Bryan's method is, after 25 years, still the {\it de facto} standard for inversion problems in condensed matter physics. One of its most attractive features is its speed: a few seconds on a laptop usually suffice to get a reasonable answer provided good starting parameters (for the grid, default model, and the range of $\alpha$) have been found. Nevertheless, the obtained answer (including the {\it a posteriori} probability distributions) should always be carefully checked.

A major issue is that the solution may strongly depend on the default model. (Note that the error bars which Ref.~\onlinecite{MaxEnt2} calculates, are conditional on the default model and do hence not reflect variability with respect to different default models.)  A practitioner usually wants to explore different (classes of) default models in order to get an idea of the robustness of the obtained answer, and sometimes to examine if lower values of $\chi^2$ can be found for other solutions which are equally smooth. In this regard, all these solutions are reminiscent of basic solutions discussed above, but the probability density of solutions is different due to the difference in protocols:
in the spirit of MEM one does not want default models that are too similar or default models that are too close to the obtained answer (an iteration where the new default model is the answer from a previous run, is considered a self-defeating strategy). This raises an important question of what strategy should be used to produce a representative set of basic solutions within MEM. One possibility is stochastic exploration of the configuration space of default models.

\section{Performance tests}
\label{sec:tests}

We perform blind tests of our method for two different kernels. The function $g(\tau_n)$ was
prepared from equation (\ref{eq:ac0}) and uncorrelated Gaussian noise was afterwards added to $g(\tau_n)$.

\subsection{Resolving the width of the high-energy peak}
\label{subsec:twopeak}

In this subsection, we assume that the spectral function $A(z)$ is identically equal to zero at
$z<0$, non-negative at $z>0$, and the kernel is $K(\tau_n,z) = {\rm e}^{-z \tau_n}$, see Eq.~(\ref{eq:spectral}).

\begin{figure}[htb]
\includegraphics[scale=0.38,width=0.90\columnwidth]{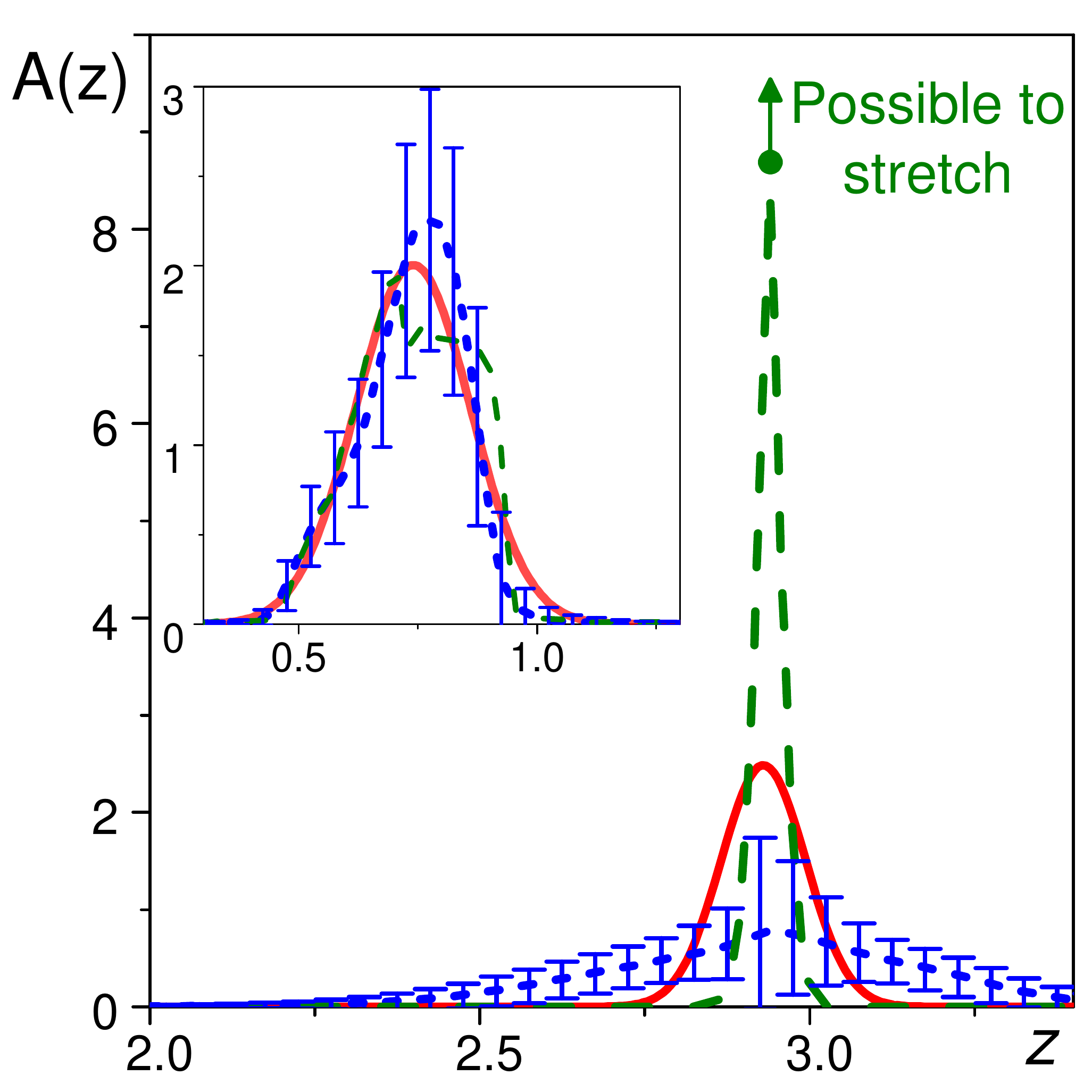}
\caption{(Color online.)
Results for test 1, featuring two peaks of finite width with noise level
$10^{-3}$. Shown is the comparison between the actual spectrum (red solid line), the smooth SOCC
spectrum (blue short-dashed line), and the pulled-up high-energy
peak SOCC solution (green dashed line). The error bars for the smooth SOCC spectrum $\{\sigma_m\}$ are determined from Eq.~(\ref{eq:disp}).
}
\label{fig:test_1_SOM}
\end{figure}
\begin{figure}[htb]
\includegraphics[scale=0.38,width=0.90\columnwidth]{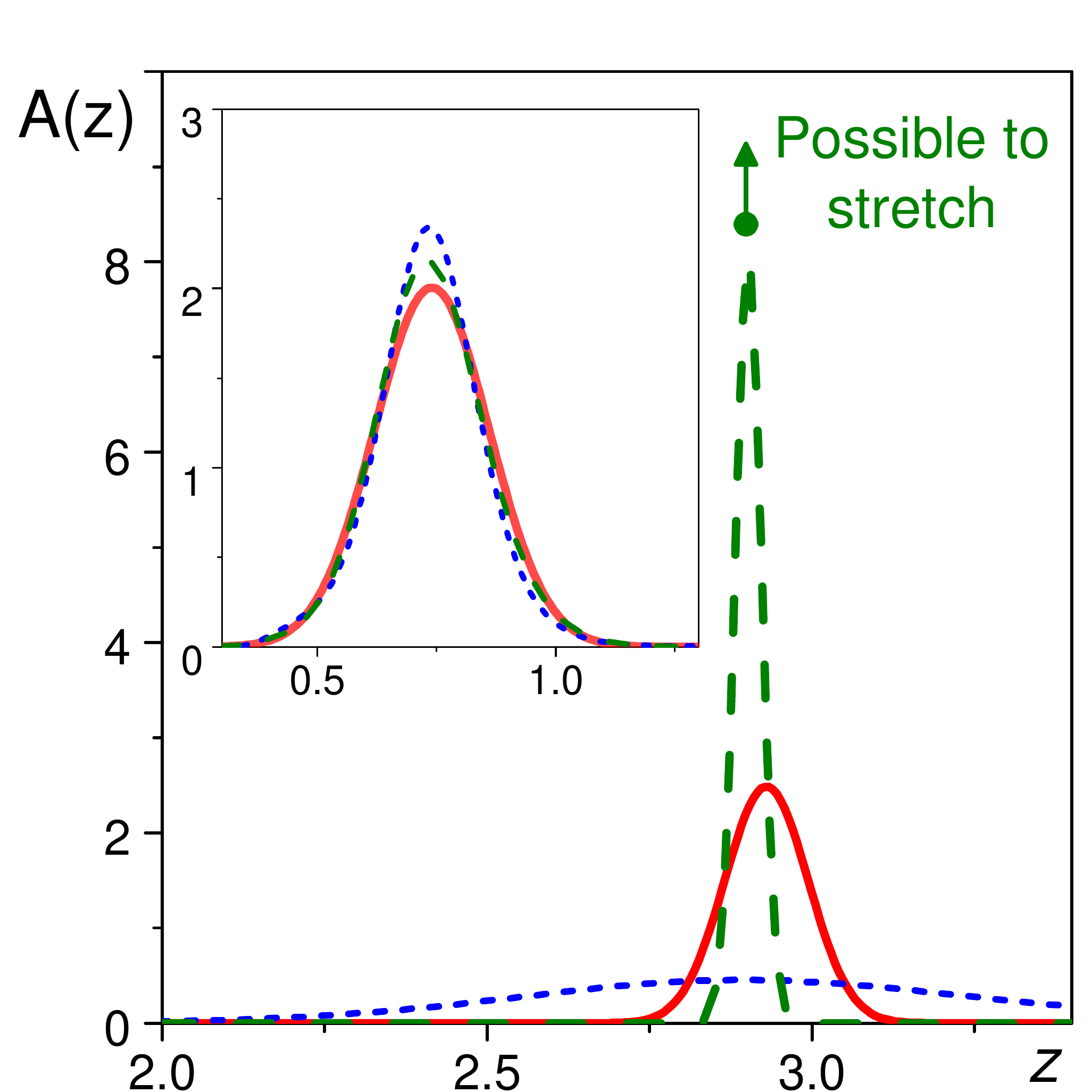}
\caption{(Color online.)
Results for test 1, featuring two peaks of finite width with noise level
$10^{-3}$. Shown is the comparison between the actual spectrum (red solid line), the
MEM spectrum with a flat default model (blue short dashed line), and the MEM spectrum with a pulled-up high-energy
peak in the default model (green dashed line).
}
\label{fig:test_1_MEM}
\end{figure}
The spectral function for test~1 and test~2 (shown in Fig.~\ref{fig:two_peaks}) contains two peaks of finite width and has the following
form (up to a normalization constant)
\begin{equation}
A(z)\, =\,
\frac{c_1}{\sigma_1}\, \exp\left\{-{(z-z_1)^2\over 2 \sigma_1^2}\right\} +
\frac{c_2}{\sigma_2}\, \exp\left\{-{(z-z_2)^2\over 2 \sigma_2^2}\right\},
\label{case_A}
\end{equation}
where $z_1=0.74$, $c_1= 0.62$, $\sigma_1=0.12$, $z_2=2.93$,
$c_2= 0.41$, and $\sigma_2=0.064$.
The spectrum is normalized to unity before adding uncorrelated Gaussian noise with relative standard deviation $\sigma=10^{-3}$
for test~1 and $\sigma=10^{-5}$ for test 2.

In the spectrum for test 3 the low-frequency peak
is not a Gaussian but a $\delta$-function with the same position and weight,
 \begin{equation}
A(z)\, =\, \sqrt{2\pi} \, c_1\, \delta (z- z_1) +
\frac{c_2}{\sigma_2}\, \exp\left\{-{(z - z_2)^2\over 2 \sigma_2^2}\right\}  \, ,
\label{case_Ad}
\end{equation}
where $z_1=0.74$, $c_1= 0.62$, $z_2=2.93$, $c_2= 0.41$, and $\sigma_2=0.064$.
The relative standard deviation of the uncorrelated Gaussian noise is $\sigma=10^{-5}$.
\begin{figure}[htb]
\includegraphics[scale=0.38,width=1.03\columnwidth]{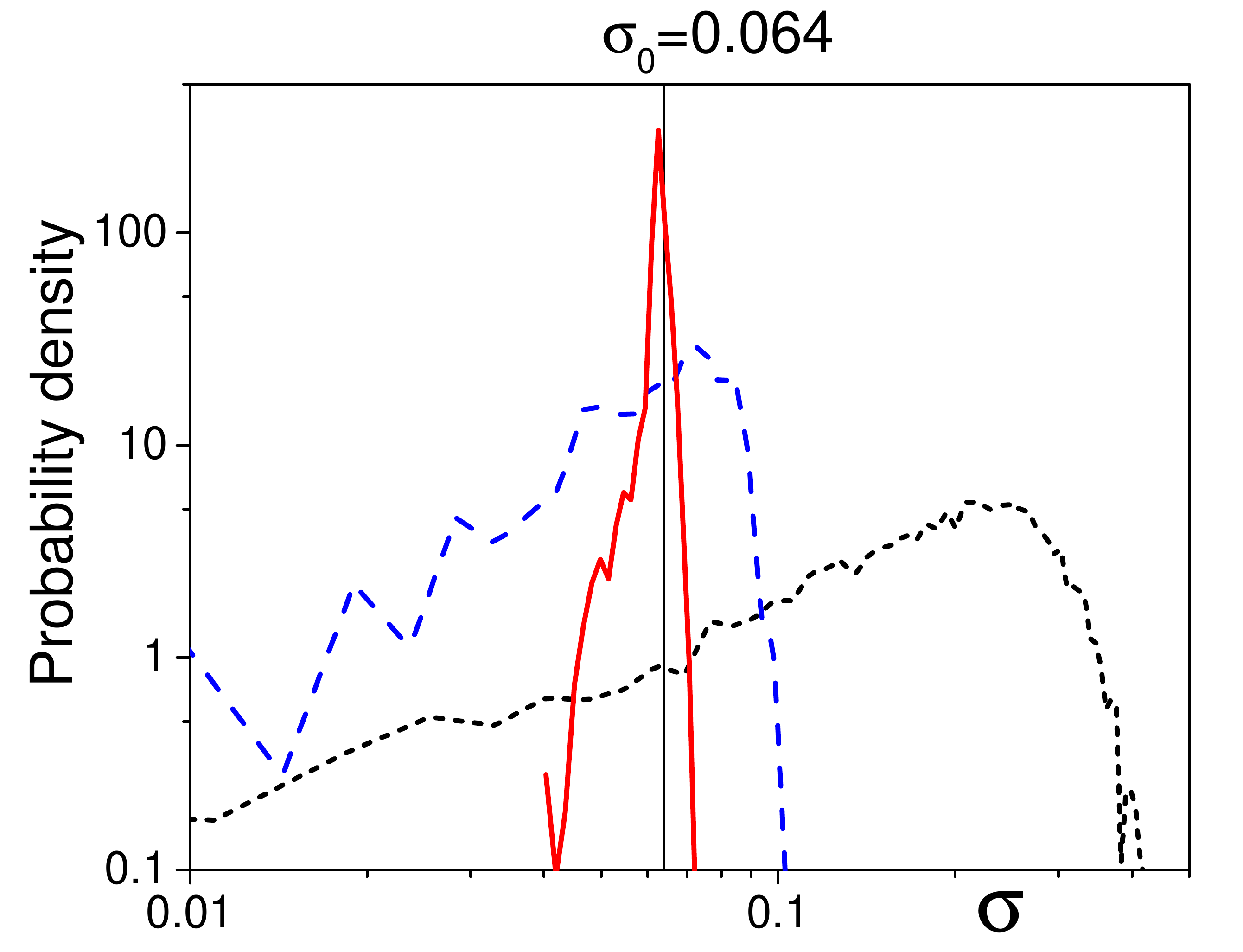}
\caption{(Color online.)
Distribution of the second moment for the
high frequency peak in test~1 (short-dashed black line), test~2
(dashed blue line), and test 3 (red solid line), among all basic solutions.
The vertical line shows the second moment $\sigma_0=0.064$ of the original spectrum.
}
\label{fig:semom}
\end{figure}

\begin{figure}[htb]
\includegraphics[scale=0.38,width=0.90\columnwidth]{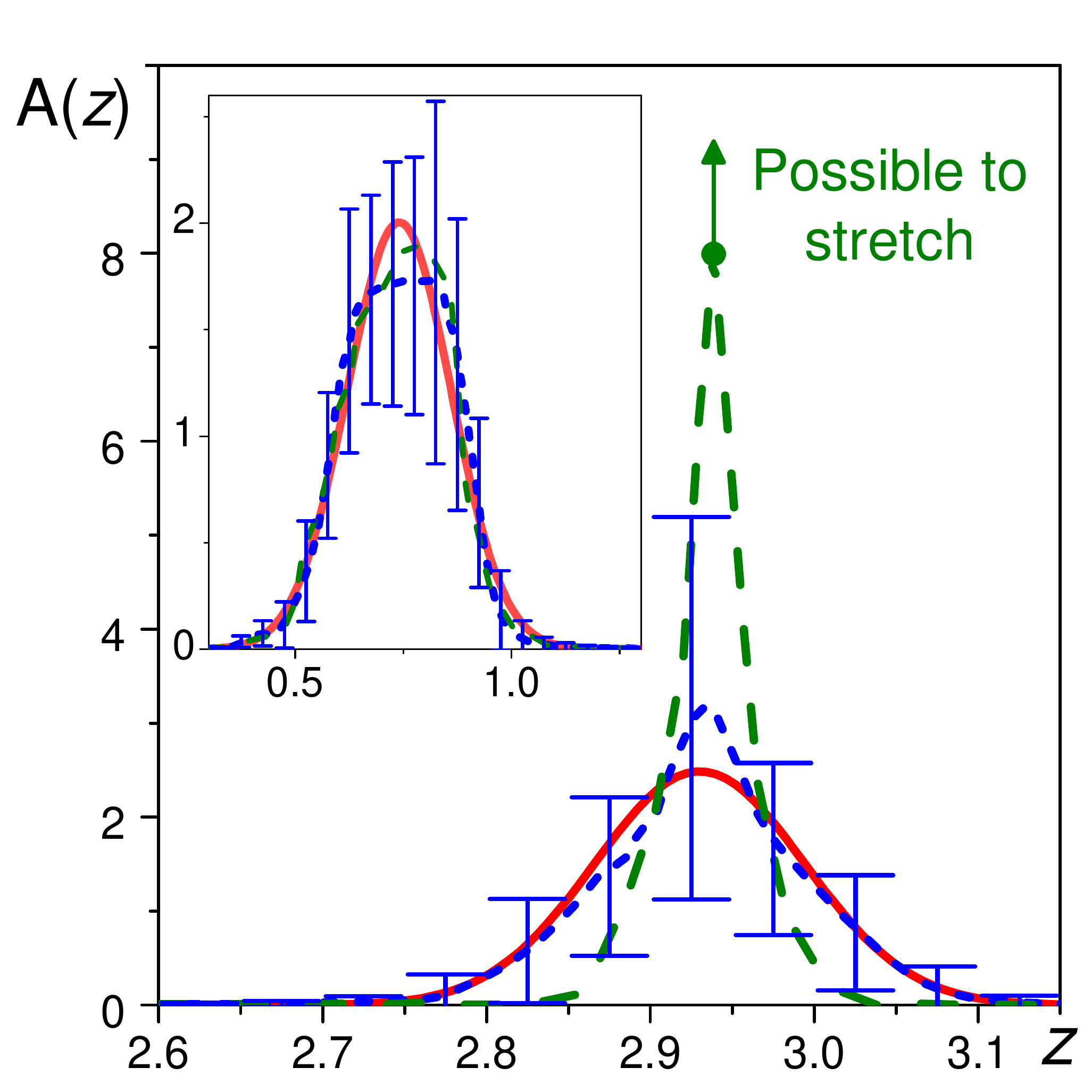}
\caption{(Color online.)
Results for test~2, featuring two peaks of finite width  with noise level
$10^{-5}$. Shown is the comparison between the actual spectrum (red solid line) and the smooth SOCC
spectrum (blue short-dashed line), the pulled-up high-energy
peak SOCC solution (green dashed line).
The error bars for smooth SOCC spectrum $\{\sigma_m\}$ determined from Eq.~(\ref{eq:disp}).
}
\label{fig:test_2_SOM}
\end{figure}
\begin{figure}[htb]
\includegraphics[scale=0.38,width=0.90\columnwidth]{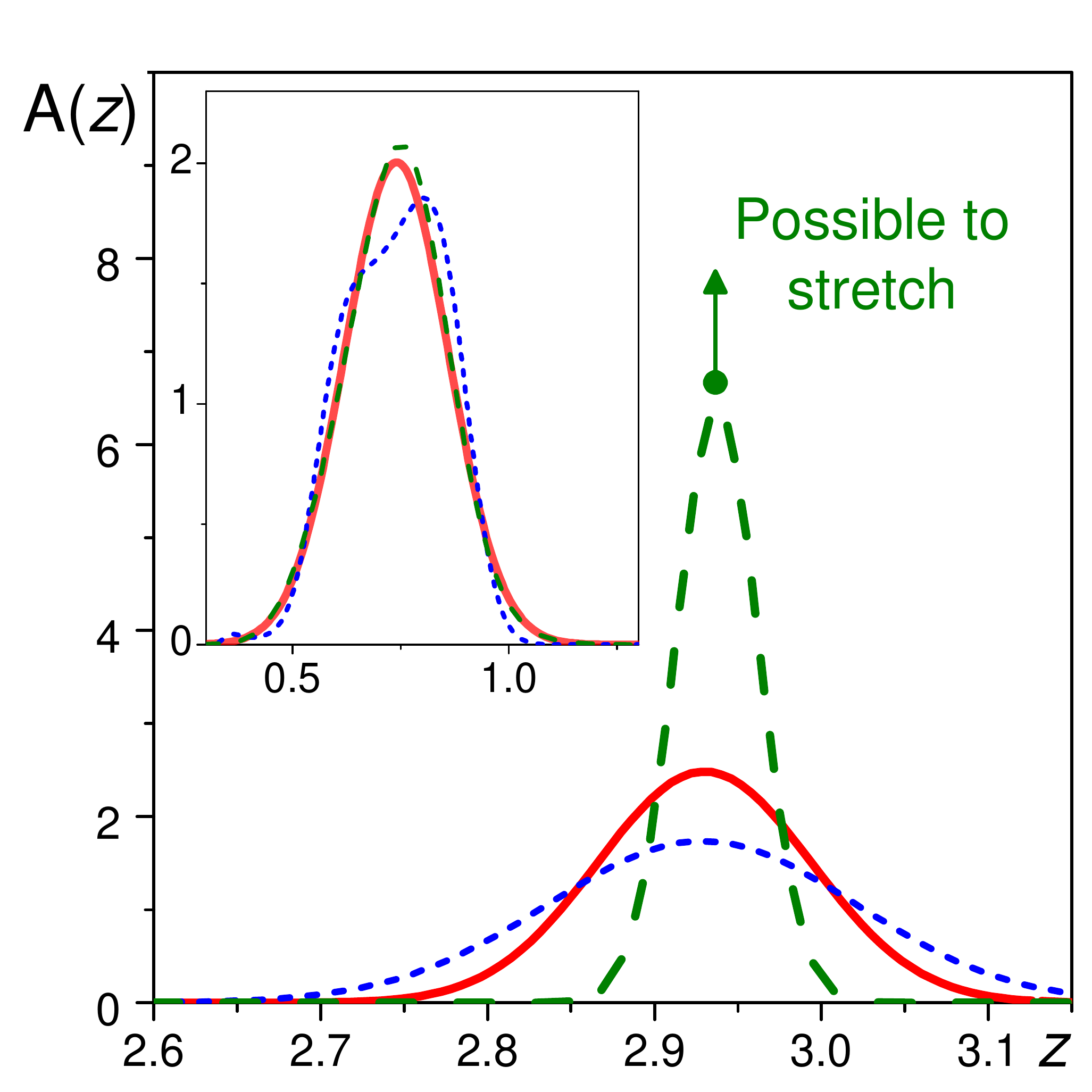}
\caption{(Color online.)
Results for test~2, featuring two peaks of finite width with noise level $10^{-5}$. Shown is the comparison between the actual spectrum (red solid line), MEM with a flat default model
(blue short-dashed line), and MEM with a pulled-up high-energy peak in the default model (green dashed line).
}
\label{fig:test_2_MEM}
\end{figure}
The challenge for NAC is to
judge whether one can resolve the width of the high-frequency peak.
To this end we consider two possible setups for MEM and SOCC.
In the standard setup we assume a flat default model for MEM and
the SOCC procedure of generating smooth solutions
as described in Sec.~\ref{subsec:final} in the absence of the default model
penalty $O_5$. To study possible deformations of the second peak, we then
introduce a narrow-peak default model in MEM and re-run the simulation, or, in the case
SOCC, we insist that the final solution goes through a much higher point at
the peak maximum. The width of the high-frequency peak is deemed impossible to resolve if one can reduce it by a factor of two, while the spectrum remains well-defined.

We note that better reproducibility of the low-frequency peak is a particular property of
the kernel (\ref{eq:spectral}).  For example, for the analytic continuation of the current-current correlation
function to the optical conductivity, the main challenge is to resolve the spectral density
at zero frequency.\cite{Mobil_15}

\begin{figure}[htb]
\includegraphics[scale=0.40,width=1.\columnwidth]{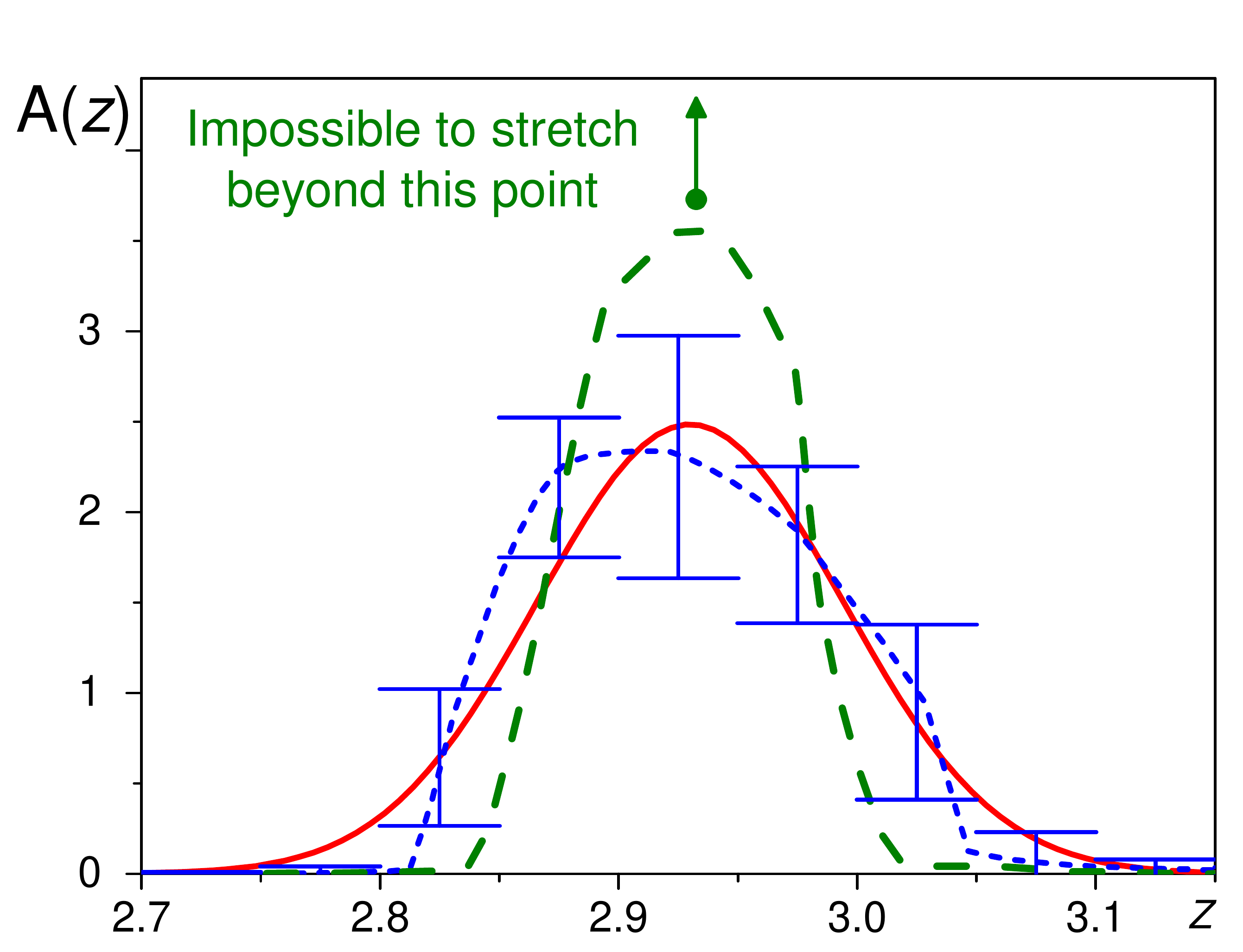}
\caption{(Color online.)
Results for test 3, characterized by a $\delta$-function at low frequency and a peak
of finite width at high frequency
with noise level $10^{-5}$. Shown is the comparison between the actual spectrum (red solid line), the smooth SOCC
spectrum (blue short-dashed line), and the maximally pulled-up high-energy
peak SOCC solution (green dashed line).
The error bars for smooth SOCC spectrum $\{\sigma_m\}$ determined from Eq.~(\ref{eq:disp}).
}
\label{fig:test_3_SOM}
\end{figure}
\begin{figure}[htb]
\includegraphics[scale=0.40,width=1.\columnwidth]{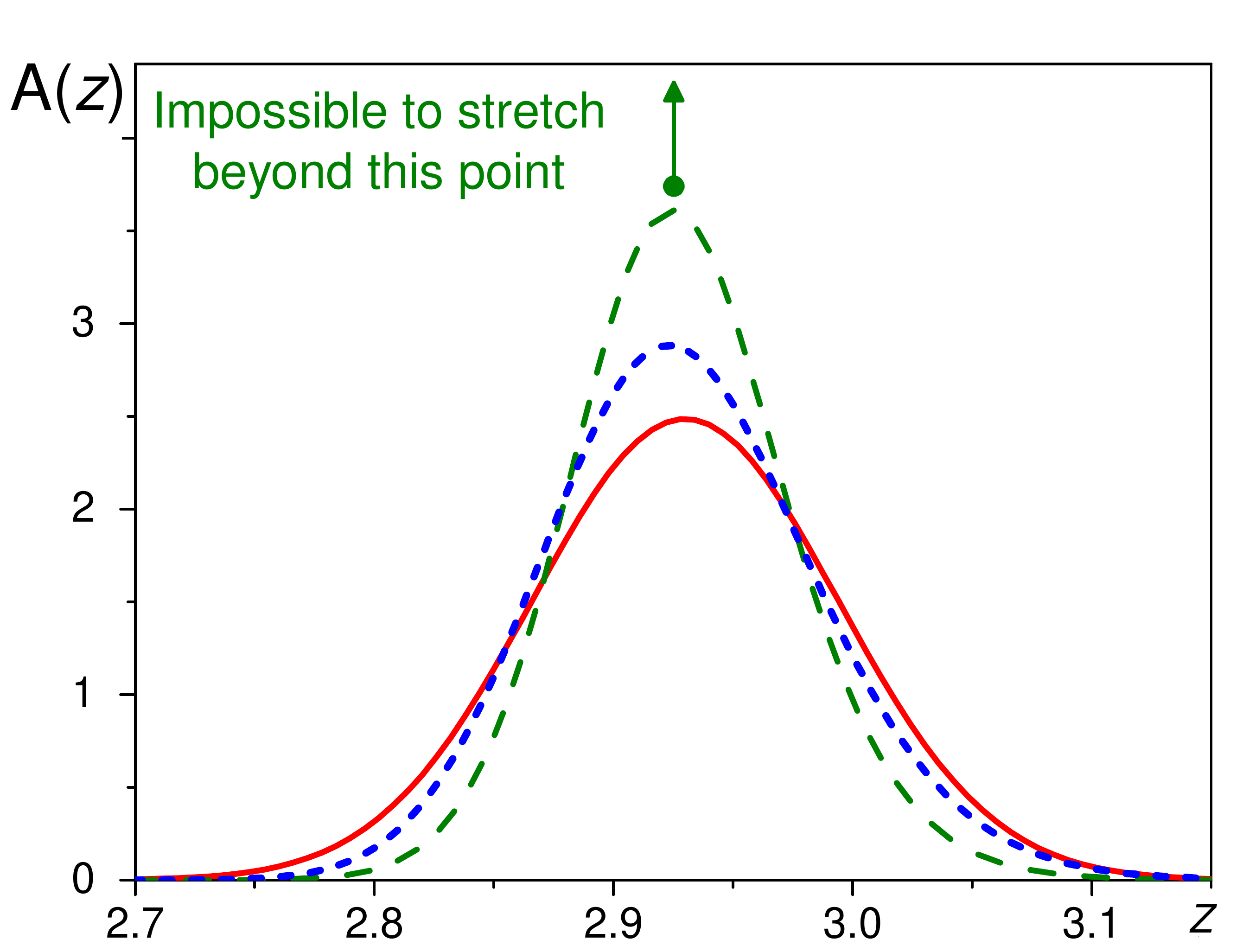}
\caption{(Color online.)
Results for test 3, characterized by a $\delta$-function at low frequency and a peak of finite width at high frequency
with noise level $10^{-5}$. Shown is the comparison between the actual
spectrum (red solid line), the MEM spectrum with a flat default model
(blue short-dashed line), and the MEM spectrum with a double Gaussian default model
where the second peak is maximally pulled up (green dashed line).
}
\label{fig:test_3_MEM}
\end{figure}

Analysis of test~1 shows that the low-energy peak can be well resolved
by both  SOCC and MEM (see the insets in Figs.~\ref{fig:test_1_SOM} and \ref{fig:test_1_MEM},
respectively). On the other hand, the high-energy peak width is severely overestimated by both methods
in the standard setup (Figs.~\ref{fig:test_1_SOM} and \ref{fig:test_1_MEM}).
However, both methods allow one to pull the high-frequency peak up at least by a factor of four above the actual spectrum.
Specifically, if the second peak in the MEM default model is set to be much narrower than the
actual one, then MEM produces an answer of the same width as this default model. Similarly, the superposition of basic SOCC solutions can be forced to have a much higher amplitude
at the second peak maximum by employing an appropriate target function.
In SOCC, one may also see direct evidence that the second peak width is questionable by
considering statistics of the second moments $\sigma$ for the high-frequency peak among all basic solutions. The corresponding distribution is presented in Fig.~\ref{fig:semom}. The probability to find a solution with a vanishing width ($\sigma \to 0$) for the high-frequency peak does not go to zero for test~1 and, hence, the imaginary-time data for $g(\tau_n)$ (within their accuracy)
do not rule out a $\delta$-function for the high-frequency peak.

Further insight is provided by test~2, which differs from test~1 only
in the noise level, which is reduced by two orders of magnitude.
One readily observes that, in contrast to test~1, the standard setups of SOCC and MEM
give a very good description of the high-frequency peak
(Figs.~\ref{fig:test_2_SOM} and \ref{fig:test_2_MEM}).
Does this mean that one can be absolutely sure that the width of the peak is finite?
The answer is no, because one can still easily pull the peak up by a factor of four.
Moreover, SOCC analysis of second moments (see Fig.~\ref{fig:semom}) demonstrates
that a $\delta$-functional shaped second peak is still a possibility, despite improved error bars.
In these examples, tighter error bars allow only to reduce the upper bound on
the width of the second peak. Much smaller error bars, which are unrealistic for Monte Carlo simulations of $g(\tau_n)$, would be required to controllably resolve
the actual width of the second peak.

Test~3 has the same Gaussian noise as test 2 but the low-frequency peak is now
replaced by a $\delta$-function, see Eq.~(\ref{case_Ad}).
This crucially changes the results. Now the high-frequency peak is well
reproduced not only in the standard setup of SOCC and MEM but also in
attempts to pull the solution up, see Figs.~\ref{fig:test_3_SOM} and \ref{fig:test_3_MEM},
respectively.
Stability of results for the second peak width is also evident in the
probability distribution for the second moment shown in Fig.~\ref{fig:semom}.
The distribution is peaked at the correct value $\sigma_2=0.064$ and is
rather narrow, indicating that a narrower peak would compromise
the error bars.

We emphasize that the success of resolving the width of the second peak in test~3 is due to a combination of two circumstances: the small width of the first peak and the high accuracy of the input data. To see this, it is instructive to consider the physical example of the Fermi polaron (see Sec.~\ref{sec:polarons}), where the width of the first peak is very small, but the accuracy of the input data is significantly lower than in test~3.

\begin{figure}[htb]
\includegraphics[scale=0.38,width=0.96\columnwidth]{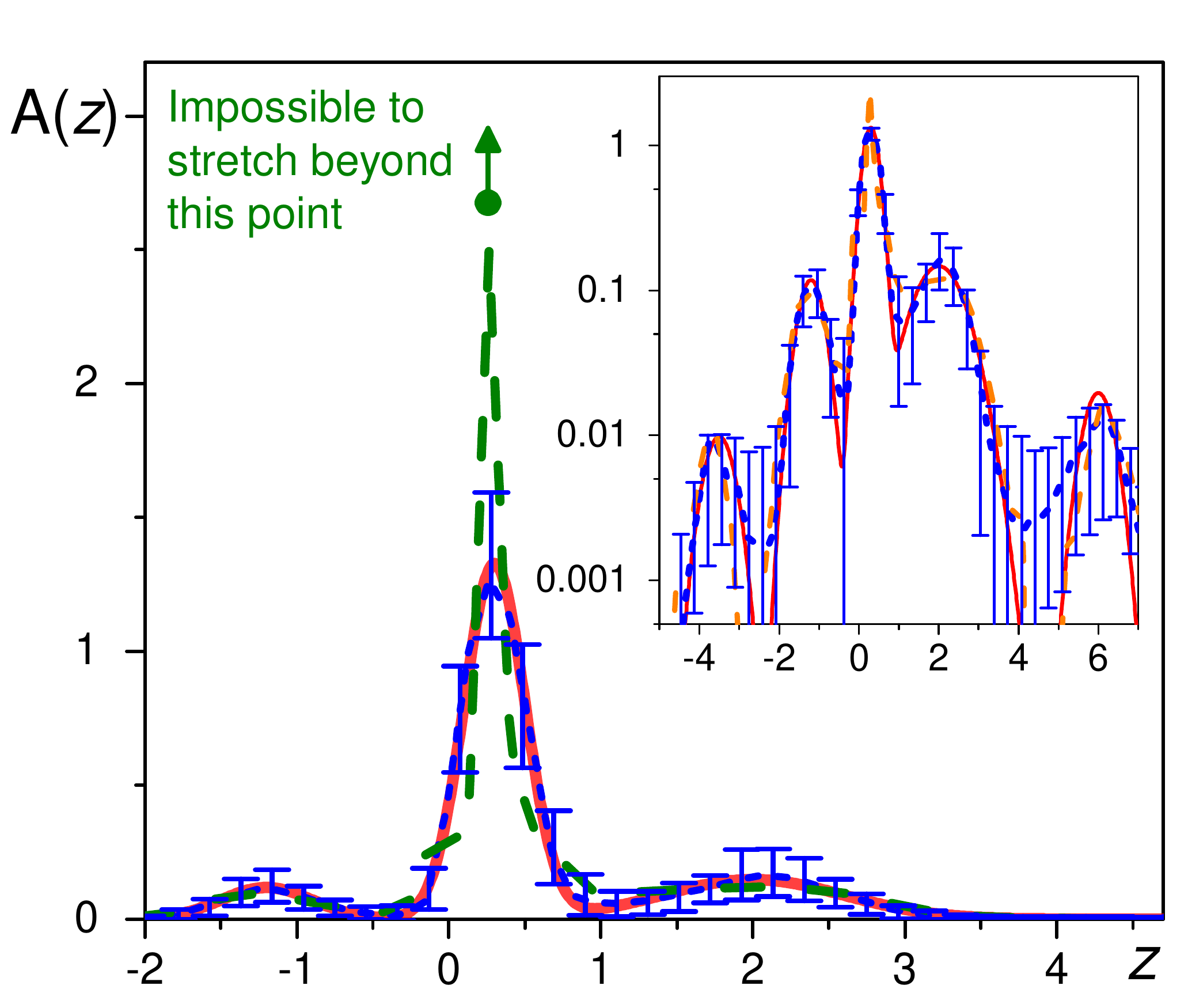}
\caption{(Color online.)
Results for test~4 with a Fermi distribution kernel and noise level
$10^{-5}$. Shown is the comparison between the actual spectrum (red solid line), the smooth SOCC
spectrum (blue short-dashed line), and the maximally pulled up central
peak SOCC solution (dashed green line).
The logarithmic plot in the inset highlights the comparison of low-intensity features.
The error bars for the smooth SOCC spectrum $\{\sigma_m\}$ are determined from Eq.~(\ref{eq:disp}).
}
\label{fig:fermi_SOM}
\end{figure}
\begin{figure}[htb]
\includegraphics[scale=0.38,width=0.96\columnwidth]{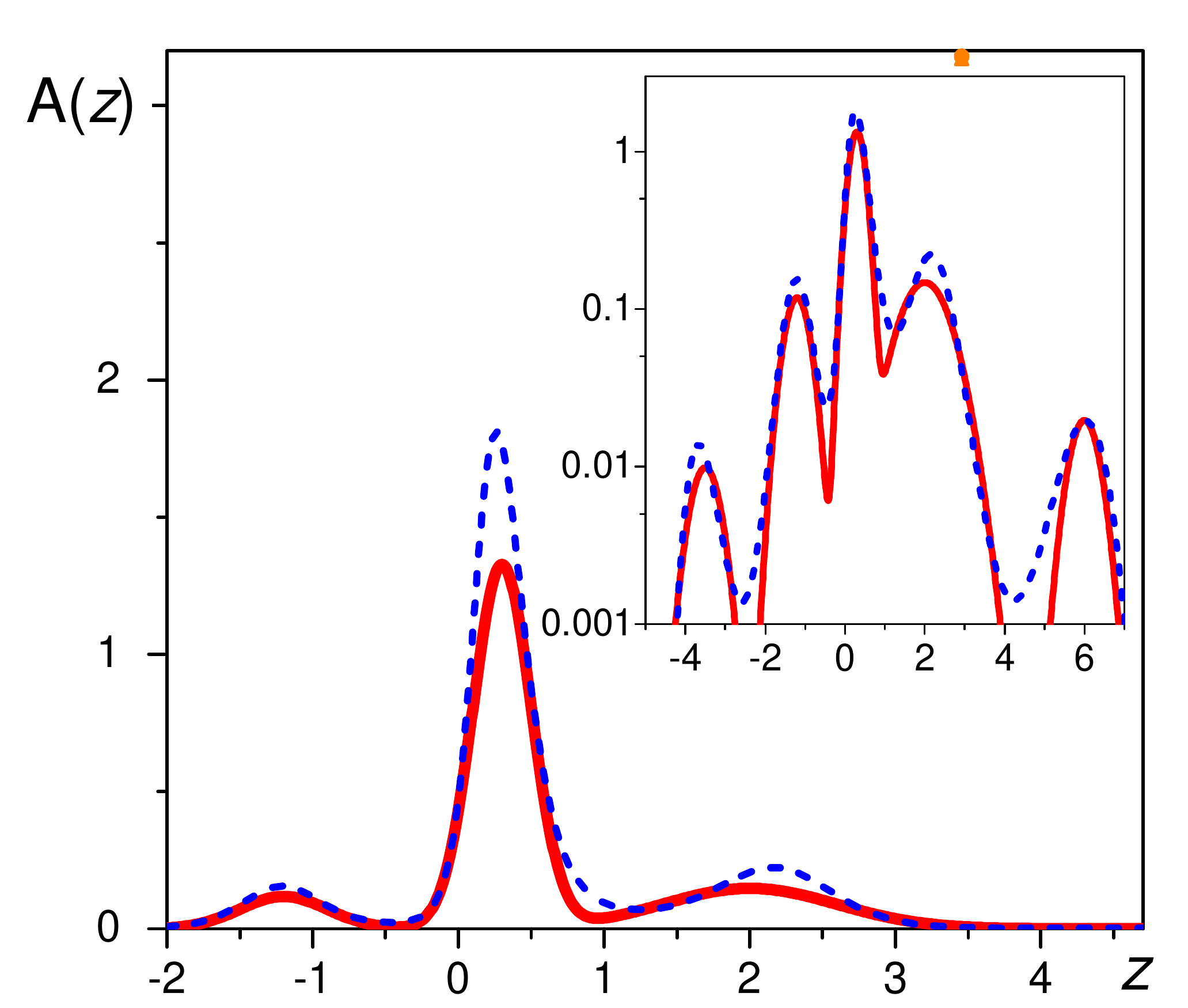}
\caption{(Color online.)
Results for test~4 with a Fermi distribution kernel and noise level
$10^{-5}$. Shown is the comparison between the actual spectrum (red solid line) and
the MEM spectrum in the default setup (blue dashed line).
}
\label{fig:fermi_MEM}
\end{figure}

\subsection{Fermi distribution kernel}
\label{subse:fermi}

Test~4 analyzes the possibility of resolving spectral densities at the Fermi level
from the analytic continuation of $g(\tau)$. Here, the spectrum $A(z)$ is defined
in the range $-\infty < z < \infty$ and the kernel is
$K(\tau_n,z) = \exp\{ -z \tau_n \} / (1+\exp\{ -z\beta \} )$ with $\beta=6$.
The uncorrelated Gaussian noise is added at the $10^{-5}$ level.
One can see in Figs.~\ref{fig:fermi_SOM} and \ref{fig:fermi_MEM}
that both SOCC and MEM give a good description of the spectral function
in the vicinity of the chemical potential (at $z=0$). Also, the height of the middle peak
at near zero frequency cannot be significantly pulled up without distorting the rest
of spectrum. Specifically for MEM, trying different default models with one, two, or three Gaussian peaks did not improve the answer; in all cases
trying to find narrower peaks resulted in secondary oscillations reminiscent of numerical instabilities. We conclude that the fermionic spectrum can be restored for the given
parameters with high quality.

\section{Application of SOCC to the Fermi polaron problem}
\label{sec:polarons}

We now test the SOCC method on a physical system---the resonant Fermi polaron (a spin-down fermion in a sea of noninteracting spin-up fermions)\cite{polaronreview, carlosreview}; here in three dimensions and for equal mass of spin-up and spin-down particles. The coupling between the polaron and its environment is characterized by a single dimensionless parameter $k_Fa$, where $k_F$ is the Fermi wave vector and $a$ the s-wave scattering length. Here we examine a typical situation at $k_Fa=0.8$ when the polaron state at zero momentum is metastable but has a very long relaxation time, implying that the lowest peak in the polaron spectral function is a sharp resonance nearly indistinguishable from a $\delta$-function.

The imaginary-time polaron Green's function at zero temperature and zero momentum, $g_n=g(\tau_n)$, was obtained with diagrammatic Monte Carlo, see Refs.~\onlinecite{SOM,polaron1,new}. We are able to achieve very high precision in our results for $g(\tau)$ with a relative error as low as $\mathcal{O}(10^{-7}-10^{-9})$ at $\tau$ close to zero, $\mathcal{O}(10^{-4}-10^{-3})$ around $\tau=1/\varepsilon_F$ (where $\varepsilon_F$ is the Fermi energy of spin-up fermions) and a few percent at the largest $\tau$ considered for the analytic continuation. The kernel at zero temperature is $K(\tau_n,z)=e^{-z\tau_n}$.

The polaron spectral function features two peaks. The position and weight of the first polaron
peak are fixed with high accuracy by the asymptotic decay of the Green's function, $-g(\tau\rightarrow\infty)\rightarrow Z_1e^{-E_1\tau}$. Our data at
large $\varepsilon_F \tau_n$ can be fitted to a single exponential (within error bars)
indicating that the polaron remains a well-defined quasi-particle in this parameter range.
The particle-hole continuum emerges at higher frequencies as a second broad peak. A key
question we want to address here is its spectral width. This has been discussed
in the context of the repulsive polaron state.\cite{parish, Kamikado, polaronfrg, repulsivepolaron, grimmexprepulsive} In order for it to qualify to be a well-defined quasiparticle, the peak width needs to be sufficiently narrow (much smaller than the Fermi energy, corresponding to a sufficiently long life time). Thus resolving the width accurately is very important to correctly interpret this spectrum. Note that this spectrum has the same general features as the spectrum for test 1 in the previous section.
\begin{figure}[htb]
\includegraphics[scale=0.38,width=1.00\columnwidth]{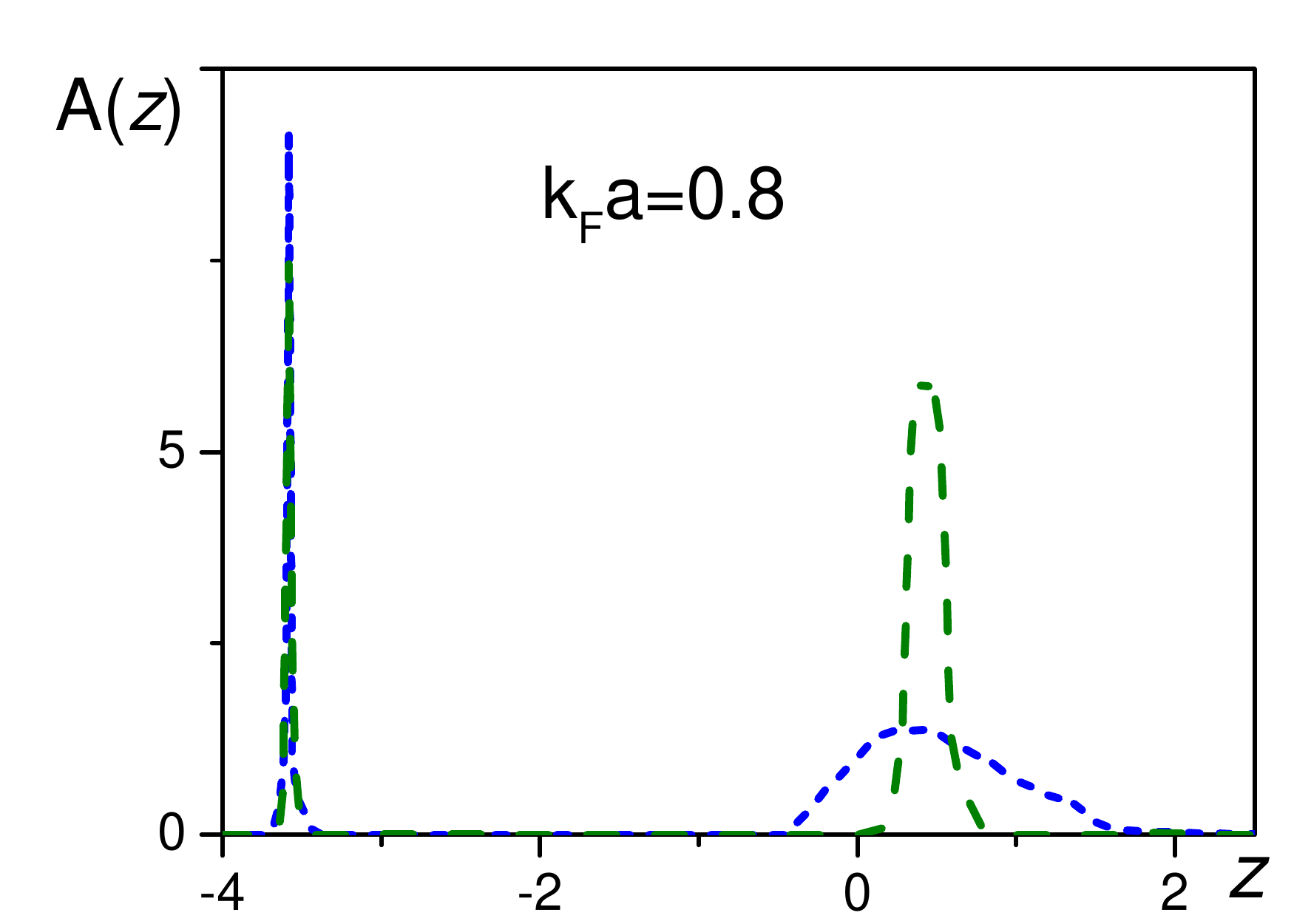}
\caption{(Color online.)
Spectral density of the resonant Fermi polaron for $k_Fa=0.8$ at zero momentum and zero temperature. The smooth SOCC spectrum
with the second peak dispersion $\sigma_2=0.48$ is shown by the blue short dashed line. However,
a much narrower solution for the second peak with $\sigma_2=0.12$ (green dashed line)
can also be obtained from the same set of basic solutions.
}
\label{fig:repulsive}
\end{figure}

When SOCC is used to produce a smooth solution the second peak emerges as a broad
spectral feature. If this was indeed the case, the metastable repulsive polaron picture would be inapplicable
for $k_Fa=0.8$. However, the same set of basic solutions can be optimized to have a much narrower peak, see Fig.~\ref{fig:repulsive}, implying that a well-defined repulsive polaron quasi-particle cannot be ruled out. Given that the second peak dispersion can be reduced by a factor of four without compromising the accuracy of the final solution, we have to conclude that
the quality of the input data is insufficient to determine the actual width.

\section{Conclusions}
\label{sec:conclusions}

The most challenging aspect of numerical analytic continuation is not the algorithm of finding a stable (smooth) solution consistent with the input data, but the protocol of assessing its accuracy and unambiguity. We have implemented such a protocol based on the method of stochastic optimization with consistent constraints and demonstrated how a similar strategy can be followed with the maximum entropy method by exploring the space of default models. Irrespective of the method, the procedure has to deal with either integrals of the spectral function (rather than the function itself) and/or certain {\it a priori} and {\it a posteriori} constraints consistent with the error bars on the input data.

It is important to distinguish between two cases. The first (simplest) case is when all physically meaningful solutions do not differ substantially, upon possible smearing of unimportant (below the resolution) fine details of the otherwise smooth spectral function. The second case, exemplified by the spectral function in Fig.~\ref{fig:two_peaks}, is when a piece of important physical information is inevitably lost. In the first case, a reasonable  characterization of uncertainties can be achieved by coarse-graining, like, e.g., Eq.~(\ref{eq:ac1}). In the second case, one has to employ a more elaborate approach to reveal the different possible physical solutions that do not compromise the error bars of the input data.

Much of our attention has been paid to the protocol of treating the second case. We have shown how it can be handled with SOCC and modified MEM. With MEM one has to explore various default
models and resulting solutions that remain consistent with input error bars. A useful feature of the SOCC approach is that such an analysis---and, more generally, the application of all possible consistent constraints---can be implemented at the post-processing stage using a representative set of ``basic" solutions generated by the stochastic-optimization protocol. The linearity of the problem (\ref{eq:ac0}) is crucial here, as it guarantees that any superposition of basic solutions with non-negative weights is also a solution to Eq.~(\ref{eq:ac0}) within the same or better level of accuracy. Even if the superposition coefficients are allowed to be negative, the procedure typically keeps the accuracy of the final solution at the level of basic solutions. This allows one to implement consistent constraints by choosing the superposition coefficients to minimize the corresponding objective function.

{\it Acknowledgements.}  This work was supported by the Simons Collaboration
on the Many Electron Problem, the National Science Foundation under
the grant PHY-1314735, the MURI Program ``New Quantum Phases of Matter" from AFOSR, and FP7/ERC starting grant No. 306897.
A.S.M. is supported by the ImPACT Program of the Council for Science, Technology and Innovation
(Cabinet Office, Government of Japan). Our maximum entropy implementation builds
on the ALPS implementation.\cite{ALPS2}



\begin{thebibliography}{99}

\bibitem{SOM}
A. S. Mishchenko, N. V. Prokof'ev, A. Sakamoto, and B.~V.~Svistunov, Phys. Rev. B {\bf 62}, 6317 (2000).

\bibitem{Julich} A. S. Mishchenko, in Verlag des Forschungszentrum J\"{u}lich, 2012. - 978-3-89336-796-2, edited by E. Pavarini, E. Koch, F. Anders, and M. Jarrell (2012).

\bibitem{CC} N. Prokof'ev and B. Svistunov,
JETP Lett. {\bf 97}, 747 (2013).

\bibitem{MaxEnt1}
R. N. Silver, D. S. Sivia, and J. E. Gubernatis,
Phys. Rev. B {\bf 41}, 2380 (1990).

\bibitem{MaxEnt1a}
R. Bryan,
Eur. Biophys. J. {\bf 18}, 165 (1990).

\bibitem{MaxEnt1b}
J. E. Gubernatis, M. Jarrell, R. N. Silver, and D. S. Sivia,
Phys. Rev. B {\bf 44}, 6011 (1991).

\bibitem{MaxEnt2}
M. Jarrell and J. E. Gubernatis,
Phys. Rep, {\bf 269}, 133 (1996).

\bibitem{MaxEnt3}
S. Fuchs, T. Pruschke, and M. Jarrell, Phys. Rev. E {\bf 81}, 056701 (2010);
A. Dirks, P. Werner, M. Jarrell, and T. Pruschke, Phys. Rev. E {\bf 82}, 026701 (2010).

\bibitem{MaxEnt5}
O. Gunnarsson, M. W. Haverkort, and G. Sangiovanni,
Phys. Rev. B {\bf 81}, 155107 (2010); {\it ibid} Phys. Rev. B {\bf 82}, 165125 (2010).

\bibitem{polaronexp1}
A. Schirotzek, C.-H. Wu, A. Sommer, and M. W. Zwierlein, Phys. Rev. Lett. {\bf 102}, 230402 (2009).

\bibitem{grimmexprepulsive}
C. Kohstall, M. Zaccanti, M. Jag, A. Trenkwalder, P. Massignan, G. M. Bruun, F. Schreck, and R. Grimm,
Nature {\bf 485}, 615618 (2012).

\bibitem{polaronexp3}
M. Koschorreck, D. Pertot, E. Vogt, B. Fr\"{o}hlich, M. Feld, and M. K\"{o}hl, Nature {\bf 485}, 619622 (2012).

\bibitem{new}
O. Goulko, A. S. Mishchenko, N. Prokof'ev, and B Svistunov, arXiv:1603.06963.

\bibitem{parish}
M. Parish, and J. Levinsen, arXiv:1608.00864.

\bibitem{Kamikado}
K. Kamikado, T. Kanazawa, and S. Uchino, arXiv:1606.03721.

\bibitem{polaronfrg}
R. Schmidt and T. Enss, Phys. Rev. A {\bf 83}, 063620 (2011).

\bibitem{repulsivepolaron}
P. Massignan, and G. M. Bruun, Eur. Phys. J. D {\bf 65}, 83 (2011).

\bibitem{Maier_note} Such a statistical treatment of multiple independent solutions was first suggested in the fast modification of SOM method, see Ref.~\onlinecite{Maier}

\bibitem{Ti1}
A. N.~Tikhonoff, Dokladyu Akademii Nauk SSSR, {\bf 39}, 195 (1943); {\it ibid} {\bf 151}, 501 (1963)
[Soviet Mathematics {\bf 4}, 1035 (1963)].

\bibitem{Ph}
D. L.~Phillips, J.~ACM {\bf 9}, 84 (1962).

\bibitem{Tikh-Arse}
A. N.~Tikhonoff and V.Y.~Arsenin,
{\it Solutions of Ill-posed Problems}
(Winston \& Sons, Washington, 1977).

\bibitem{Tikh_etal}
A. N. Tikhonov, A. Goncharsky, V. V. Stepanov, and A. Yagola,
{\it Numerical Methods for the Solution of Ill-posed Problems}
(Springer-Science+Business Media, B.V., Moscow, 1995).

\bibitem{svd}
C. E. Creffield, E. G. Klepfish, E. R. Pike, and S. Sarkar,
Phys. Rev. Lett. {\bf 75}, 517 (1995).

\bibitem{nnls}
C. L. Lawson and R. J. Hanson,
{\it Solving Least Squares Problems},
Society for Industrial and Applied Mathematics, Philadelphia,
1995.

\bibitem{sr}
I. S. Krivenko and A. N. Rubtsov, JETP Lett. {\bf 94}, 768 (2012).

\bibitem{pade}
J. Sch\.{o}tt, I. L. M. Locht, E. Lundin, O. Gr\r{a}n\`{a}s, O. Eriksson, and I. Di Marco,
Phys. Rev. B {\bf 93}, 075104 (2016).

\bibitem{pade2}
J. Sch\.{o}tt, E. G. C. P. van Loon, I. L. M. Locht, M. Katsnelson, and I. Di Marco,
arXiv:cond-mat/1607.04212.

\bibitem{Sand}
A. W. Sandvik,
Phys. Rev. B {\bf 57}, 10287 (1998).

\bibitem{Vafayi}
K. Vafayi and O. Gunnarsson, Phys. Rev. B {\bf 76}, 035115 (2007).

\bibitem{Beach}
K. S. D. Beach, arXiv:cond-mat/0403055.

\bibitem{Sandvik2015}
A. W. Sandvik, arXiv:1502.06066.

\bibitem{Maier} T. A. Maier, private communication.

\bibitem{Mobil_15} A. S. Mishchenko, N. Nagaosa, G. De Filippis, A. de Candia, and V. Cataudella, Phys. Rev. Lett. {\bf 114}, 146401 (2015).

\bibitem{polaronreview}
P. Massignan, M. Zaccanti, and  G. M. Bruun, Rep. Prog. Phys. {\bf 77}, 034401 (2014).

\bibitem{carlosreview}
Z. Lan and C. Lobo, J. Indian I. Sci. {\bf 94}, 179 (2014).

\bibitem{polaron1}
N. Prokof'ev and B. Svistunov, Phys. Rev. B {\bf 77}, 020408 (2008); ibid Phys. Rev. B {\bf 77}, 125101 (2008).

\bibitem{ALPS2}
B. Bauer {\it et al.}, Journal of Statistical Mechanics: Theory and Experiment, P05001 (2011).

\end{thebibliography}
\end{document}